\documentclass[twocolumn,article,prl,superscriptaddress,letterpaper,floatfix]{revtex4}
\usepackage[english]{babel}
\usepackage[T1]{fontenc}
\usepackage[colorlinks=true,linkcolor=blue,citecolor=blue,urlcolor=blue]{hyperref}

\usepackage{graphicx}
\usepackage{bm,bbm}
\usepackage{amsmath,amssymb}%
\usepackage{epstopdf}
\usepackage{color}
\usepackage{isotope}
\usepackage{bbding}

\usepackage[capitalize]{cleveref}

\usepackage{multirow,booktabs,dcolumn}

\newcommand{\prel}{\ifmmode p_{rel} \else $p_{rel}$ \fi}
\newcommand{\pcm}{\ifmmode p_{cm} \else $p_{cm}$ \fi}
\newcommand*{\MIT }{Massachusetts Institute of Technology, Cambridge, Massachusetts 02139, USA}
\newcommand*{\FRIB}{Facility for Rare Isotope Beams, Michigan State University, East Lansing, Michigan 48824, USA}
\newcommand*{\LANL}{Theoretical Division, Los Alamos National Laboratory, Los Alamos, New Mexico 87545, USA}
\newcommand*{\ODU}{Old Dominion University, Norfolk, Virginia 23529, USA}
\newcommand*{\JLAB}{Thomas Jefferson National Accelerator Facility, Newport News, Virginia 23606, USA}
\newcommand*{\TAU }{School of Physics and Astronomy, Tel Aviv University, Tel Aviv 69978, Israel}
\newcommand*{\HUJI}{Racah Institute of Physics, The Hebrew University, Jerusalem 91904, Israel}
\newcommand*{\ANL}{Physics Division, Argonne National Laboratory, Lemont, Illinois 60439, USA}

\begin{document}

\title{Scale and Scheme Independence and Position-Momentum Equivalence of Nuclear Short-Range Correlations}

\author{R. Cruz-Torres}
\affiliation{\MIT}

\author{D. Lonardoni}
\affiliation{\FRIB}
\affiliation{\LANL}

\author{R. Weiss}
\affiliation{\HUJI}

\author{N. Barnea}
\affiliation{\HUJI}

\author{D. W. Higinbotham}
\affiliation{\JLAB}

\author{E. Piasetzky}
\affiliation{\TAU}

\author{A. Schmidt}
\affiliation{\MIT}

\author{L. B. Weinstein}
\affiliation{\ODU}

\author{R. B. Wiringa}
\affiliation{\ANL}

\author{O. Hen}
\email[Contact author: ]{hen@mit.edu}
\affiliation{\MIT}


\begin{abstract}
Ab-initio Quantum Monte Carlo (QMC) calculations of nuclei from deuterium to \isotope[40]{Ca}, obtained using four different phenomenological and local chiral nuclear potentials, are analyzed using the Generalized Contact Formalism (GCF). We extract spin- and isospin-dependent ``nuclear contact terms'' for each interaction in both coordinate and momentum space. The extracted contact terms, that count the number of short-range correlated (SRC) pairs with different quantum numbers, are dependent on the nuclear interaction model used in the QMC calculation. However, the ratios of contact terms for a nucleus $A$ to deuterium (for spin-1 $pn$ pairs) or to \isotope[4]{He} (for all $NN$ pairs) are independent of the nuclear interaction model and are the same for both short-distance and high-momentum pairs. This implies that the relative abundance of {\it short-range} pairs in the nucleus is a {\it long-range} (mean-field) quantity that is insensitive to the short-distance nature of the nuclear force. Measurements of exclusive $(e,e'NN)$ pair breakup processes are instead more sensitive to short-range dynamics.

\end{abstract}

\maketitle


A full theoretical description of the nuclear many-body wave function is an outstanding challenge. While mean-field approximations, such as the nuclear shell model, well describe many bulk properties of nuclei, they fail to capture the direct effects of two- and many-body correlations. Specifically, effective nuclear models struggle to describe the short-distance and high-momentum components of the nuclear many-body wave function.

Experimental and theoretical studies have shown that this part of the wave function is dominated by short-range correlations (SRCs): pairs of nucleons with large relative and individual momenta and smaller center-of-mass (c.m.) momenta, where large is measured relative to the typical nuclear Fermi momentum $k_F \approx 250\,\rm MeV/c$~\cite{Hen:2016kwk,Atti:2015eda}. At momenta just above $k_F$ $(300 \le k \le 600\,\rm MeV/c)$, SRCs are dominated by $pn$ pairs~\cite{tang03,piasetzky06,subedi08,baghdasaryan10,korover14,hen14,duer18,Duer:2018sxh}. This $pn$ dominance is due to the tensor part of the nucleon-nucleon ($NN$) interaction~\cite{schiavilla07, alvioli08, sargsian05}. SRCs in nuclei, and their specific characteristics, have implications for the internal structure of nucleons bound in nuclei~\cite{weinstein11,Hen12,Hen:2013oha,Hen:2016kwk,Schmookler:2019nvf}, neutrinoless double beta decay matrix elements~\cite{Kortelainen:2007rh,Kortelainen:2007mn,Menendez:2008jp,Simkovic:2009pp,Benhar:2014cka,Cruz-Torres:2017sjy, Wang:2019hjy}, nuclear charge radii~\cite{Miller:2018mfb}, and the nuclear symmetry energy and neutron star properties~\cite{Li:2018lpy,hen15,frankfurt08b}.

As nucleons in SRC pairs have significant spatial overlap and are far off their mass-shell, their theoretical description poses a significant challenge. Ab-initio many-body calculations using different models of the two- and three-nucleon ($3N$) interaction produce nuclear wave functions that differ significantly at high-momenta and at short-distances~\cite{Wiringa:2014,Carlson:2015,Lonardoni:2018nofk} (see \cref{Fig:UniversalFunctions} and online Supplementary Materials). This is generally referred to as ``scale and scheme dependence'', where ``scheme'' refers to the type of interaction (e.g., phenomenological or derived from Chiral Effective Field Theory, $\chi$EFT), and ``scale'' refers to the regulation cut-off inherent to EFT models. This dependence raises important questions about the model dependence of the interpretation of SRC measurements and of their implications for other phenomena.

We study the scale and scheme dependence of SRCs using the Generalized Contact Formalism (GCF)~\cite{Weiss:2015mba,Weiss:2016obx,Weiss:2018tbu} to analyze ab-initio Quantum Monte Carlo (QMC) calculations~\cite{Carlson:2015,Lynn:2019} of two-nucleon densities in nuclei from $A=2$ to $40$~\cite{Wiringa:2014,Carlson:2015,Lynn:2017,Lonardoni:2017,Lonardoni:2018prl,Lonardoni:2018prc,Lonardoni:2018nofk,Lynn:2019a2,Wiringa:qmc}. The calculations are carried out using four different $NN$+$3N$ interactions. Two interactions are the phenomenological Argonne V18 (AV18)~\cite{Wiringa:1995} + Urbana X (UX)~\cite{Pudliner:1997,Wiringa:2014} and its reduction AV4'~\cite{Wiringa:2002} + UIX$_{\rm c}$~\cite{Lynn:2019a2} which does not include a tensor operator. The other two are local $\chi$EFT interactions at next-to-next-to-leading order (N$^2$LO) with regulation scales (coordinate-space cutoffs) of $1.0$ and $1.2\,\rm fm$~\cite{Gezerlis:2013,Gezerlis:2014,Lynn:2016,Lonardoni:2018prc}. 

Here, using a common factorized framework to compare calculations employing these interactions, we learn which properties of SRCs are scale and scheme independent, and which are sensitive to the details of the nuclear interaction at short distance. We further explore the connection between nucleon pairs at small separation and at high relative momentum to obtain a consistent understanding of the nuclear many-body wave function at short distance and high momentum.

\section{Generalized Contact Formalism and Quantum Monte Carlo Calculations}

\begin{figure}[t]
\includegraphics[width=\columnwidth]{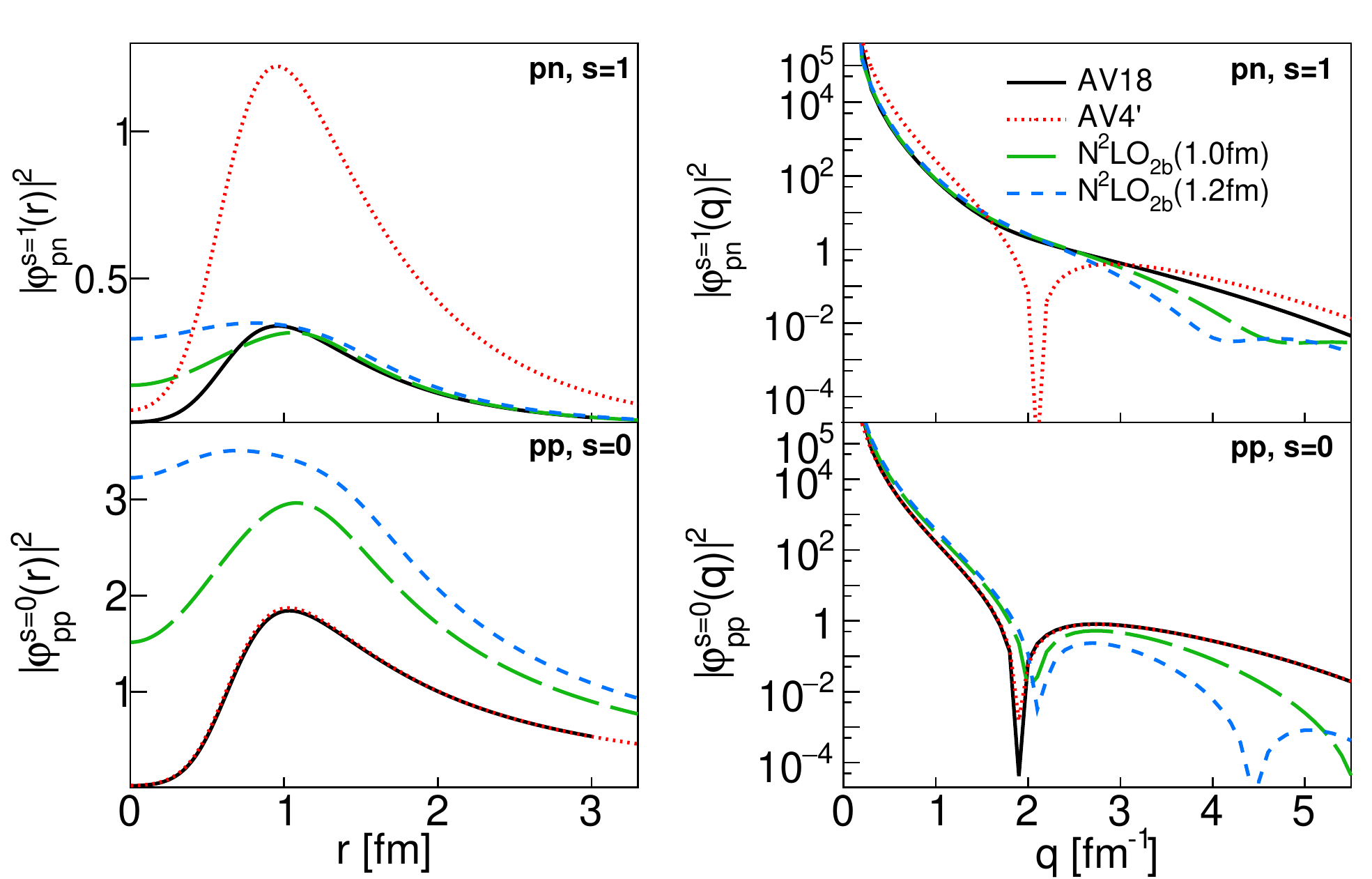}
\caption{Universal two-body functions $|\varphi_{NN}^{\alpha}|^2$ for spin-1 $pn$ (top) and spin-0 $pp$ (bottom) pairs calculated in both coordinate (left) and momentum (right) space for four different $NN$ potentials. See text for details.}
\label{Fig:UniversalFunctions}
\end{figure}

The GCF  is an effective model that provides a factorized approximation for the short-distance (small-$r$) and high-momentum (large-$k$) components of the nuclear many-body wave function. Its derivation is rooted in the scale separation between the strong relative interaction of nucleons in SRC pairs and their weaker interaction with the residual $A-2$ nuclear system~\cite{Weiss:2015mba,Weiss:2016obx,Cohen:2018gzh}. Using this scale separation, the two-nucleon density in either coordinate or momentum space (i.e., the probability of finding two nucleons with separation $r$ or relative momentum $q$) can be expressed at small separation or high momentum as~\cite{Weiss:2016obx}:
\begin{align}
	\rho_A^{NN,\alpha}(r) & = C_A^{NN, \alpha} \times |\varphi_{NN}^{\alpha}(r)|^2 , \nonumber \\
	n_A^{NN,\alpha}(q) & = C_A^{NN, \alpha} \times |\varphi_{NN}^{\alpha}(q)|^2 , 
\label{Eq1}
\end{align}
where $A$ denotes the nucleus, $NN$ denotes the nucleon pair being considered ($pn$, $pp$, $nn$), and $\alpha$ stands for the nucleon-pair quantum state (spin 0 or 1). $C_A^{NN, \alpha}$ are nucleus-dependent scaling coefficients, referred to as ``nuclear contact terms'', and $\varphi_{NN}^{\alpha}$ are universal two-body wave functions that are given by the zero-energy solution of the two-body Schr\"odinger equation for the $NN$ pair in the state $\alpha$. $\varphi_{NN}^{\alpha}$ are universal in the sense that they are nucleus independent, but they do strongly depend on the $NN$ interaction model, see \cref{Fig:UniversalFunctions}. 

While the normalizations of two-nucleon densities are well defined by the total number of nucleons in the nucleus, the individual normalizations of $C_A^{NN, \alpha}$ and $|\varphi_{NN}^{\alpha}|^2$ are not. We therefore choose to normalize $|\varphi_{NN}^{\alpha}(q)|^2$ such that its integral above $1.3\,\rm fm^{-1} (\approx k_F)$ equals unity~\cite{Weiss:2016obx}. $\varphi_{NN}^{\alpha}(r)$ is the Fourier transform of $\varphi_{NN}^{\alpha}(q)$. Thus, the normalization of one function automatically defines the normalization of the other.

We note that an important feature of the GCF is the equivalence between short distance and high momentum. This is built into \cref{Eq1} by the use of the same contact terms $C_A^{NN, \alpha}$ for both densities. 

Previous studies of the GCF~\cite{Weiss:2016obx} showed the validity of \cref{Eq1} using QMC calculations of $\rho_A^{NN,\alpha}(r)$ and $n_A^{NN,\alpha}(q)$ for $A = 2$ to 40 using the AV18+UX interaction.
More recently, the authors of Refs.~\cite{Chen:2016bde,Lynn:2019a2} analyzed QMC calculations of $\rho_A^{NN}(r)$ using the same four interactions studied here (although without separating them into different spin-isospin channels), and showed the first evidence for scale-and-scheme independence of $\rho_A^{NN}(r)/\rho_d^{NN}(r)$ ratios for short distances.

Here we extend these previous studies using new QMC calculations of two-nucleon distributions in both coordinate and momentum space, projected into different spin-isospin channels, for different nuclei and using different $NN$+$3N$ potentials (AV18+UX, AV4'+UIX$_{\rm c}$, N$^2$LO $(1.0\,\rm fm)$ and N$^2$LO $(1.2\,\rm fm)$), see \cref{Table:1}. 

\begin{table}[b]
\caption{QMC-calculated two-nucleon distributions for different nuclei and $NN$+$3N$ potentials. Checkmarks indicate calculations used in the current study. All calculations are available for both coordinate and momentum space, except for \isotope[16]{O} and \isotope[40]{Ca} with AV18 (labeled with an asterisk below), for which the UIX potential is used and results are only available in coordinate space~\cite{Lonardoni:2017}. Calculations with the N$^2$LO $(1.2\,\rm fm)$ potential for heavier systems are not considered in this work due to the large regulator artifacts found for $A\ge12$ (see Ref.~\cite{Lonardoni:2018prc}).}
\begin{ruledtabular}
\begin{tabular}{ccccc}
\isotope[A]{Z}   & AV18+UX            & AV4'+UIX$_{\rm c}$ & N$^2$LO$(1.0\rm fm)$ & N$^2$LO$(1.2\rm fm)$ \\ 
\hline                                                                                                                        
d                & \Checkmark         & \Checkmark         & \Checkmark    & \Checkmark    \\ 
\isotope[3]{H}   & \Checkmark         & \Checkmark         & \Checkmark    & \Checkmark    \\ 
\isotope[3]{He}  & \Checkmark         & \Checkmark         & \Checkmark    & \Checkmark    \\ 
\isotope[4]{He}  & \Checkmark         & \Checkmark         & \Checkmark    & \Checkmark    \\ 
\isotope[6]{Li}  & \Checkmark         & \Checkmark         & \Checkmark    & \Checkmark    \\ 
\isotope[12]{C}  & \Checkmark         & \Checkmark         & \Checkmark    & --            \\ 
\isotope[16]{O}  & \;\,\Checkmark$^*$ & \Checkmark         & \Checkmark    & --            \\ 
\isotope[40]{Ca} & \;\,\Checkmark$^*$ & \Checkmark         & --            & --            \\
\end{tabular}
\end{ruledtabular}
\label{Table:1}
\end{table}

\begin{figure*}[t]
\includegraphics[width=\columnwidth]{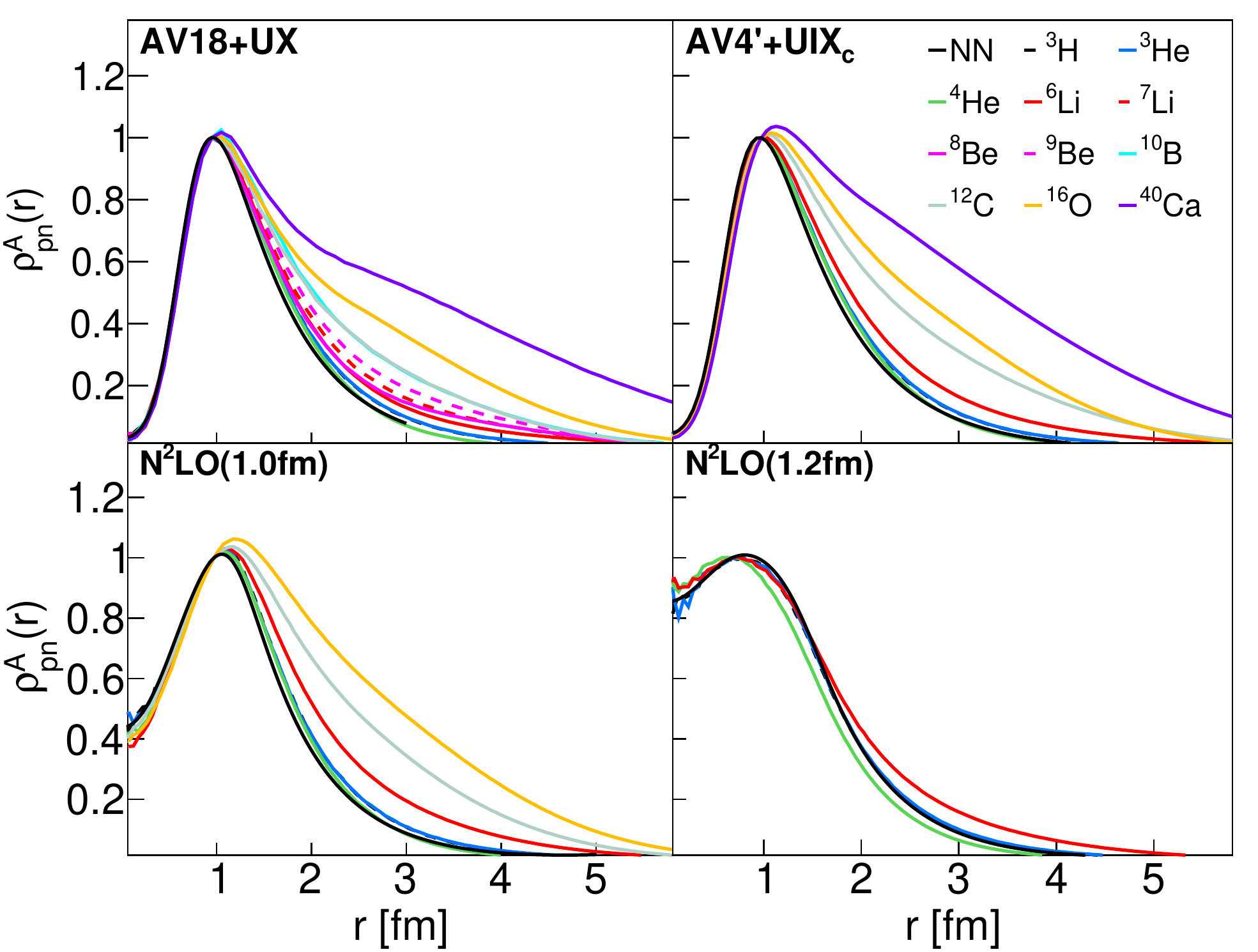}\quad
\includegraphics[width=\columnwidth]{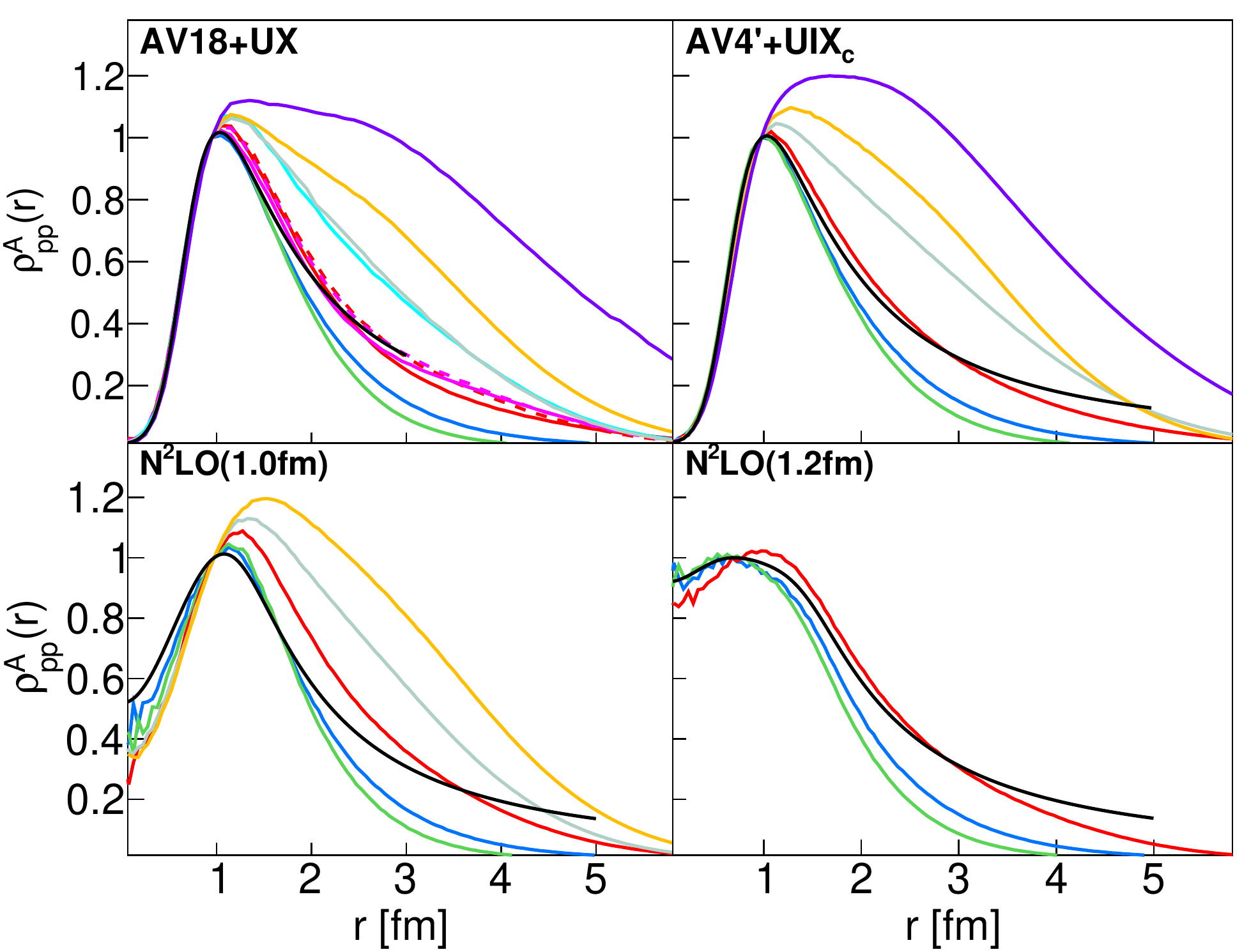}
\caption{QMC calculations of two-nucleon coordinate-space distributions for $pn$ (left) and $pp$ (right) pairs calculated using different $NN$+$3N$ potentials for different nuclei (colored lines) compared with the two-body universal function $|\varphi_{NN}^{\alpha}(r)|^2$ (solid black lines, $pn$ spin-1 on the left, $pp$ spin-0 on the right). For each interaction, QMC results are scaled to have the same value at $\sim 1\,\rm fm$ and show the same short-distance behavior for all nuclei. The difference between $NN$ distributions in the same nucleus obtained using different interactions, as shown by the four panels, indicates the scale and scheme dependence of the many-body calculations. See text for details.}
\label{Fig:DensityScaling_r}
\end{figure*}

The phenomenological AV18~\cite{Wiringa:1995} and AV4'~\cite{Wiringa:2002} potentials are ``hard interactions'', with a significant probability for nucleons to have high momentum ($k > 3$ fm$^{-1}$, see Fig. \ref{Fig:UniversalFunctions}). Their derivation is similar with AV4' being a reprojection of AV18 onto the first four channels that does not include the tensor interaction. Both potentials are supported by $3N$ forces (UX~\cite{Pudliner:1997} and the central component of UIX, UIX$_{\rm c}$~\cite{Lynn:2019}, respectively), that provide a good description of all nuclei considered in this work~\cite{Carlson:2015,Lonardoni:2017,Lynn:2019}. 

The N$^2$LO interactions are fundamentally different, as they are based on a chiral perturbation expansion up to third order with local coordinate-space regulators at distances of $1.0$ and $1.2\,\rm fm$~\cite{Gezerlis:2013,Gezerlis:2014,Lynn:2016,Lonardoni:2018prc}. These regulators make them much softer, i.e., their single-nucleon momentum distributions have much less high-momentum strength ($k > 3$ fm$^{-1}$, see Fig. \ref{Fig:UniversalFunctions}) as compared to AV18 and AV4'.

The different potentials are used to construct the fully correlated many-body wave functions used in Variational Monte Carlo (VMC) calculations. These wave functions are then propagated in imaginary-time via Diffusion Monte Carlo (DMC) techniques in order to access the true ground state of the system for a given interaction model. Quantities, such as two-nucleon distributions, are in general extrapolated from both VMC and DMC results. However, momentum-space calculations are currently available only at the VMC level~\cite{Carlson:2015,Wiringa:2014,Lonardoni:2018prc,Lonardoni:2018nofk}. Therefore, for consistency between two-nucleon distributions in coordinate and momentum space, we only use results for $\rho_A^{NN}(r)$ obtained from VMC calculations. At short distance, these densities are almost identical to the DMC and extrapolated results. We therefore use these small differences as a measure of the QMC systematic uncertainty. See online Supplementary Materials.


\begin{figure*}[t]
\includegraphics[width=\columnwidth,height=6.01cm]{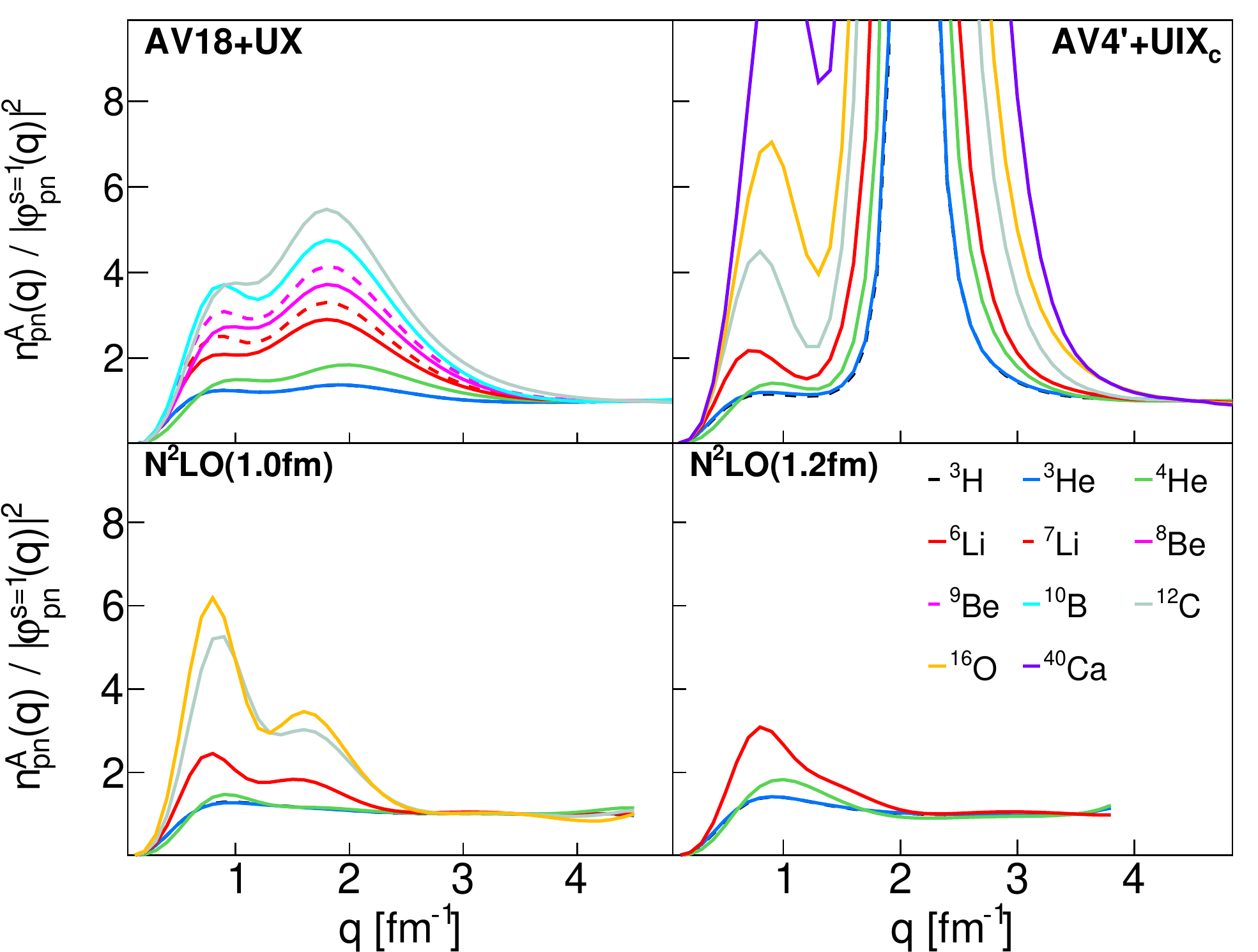}\quad
\includegraphics[width=\columnwidth,height=6.01cm]{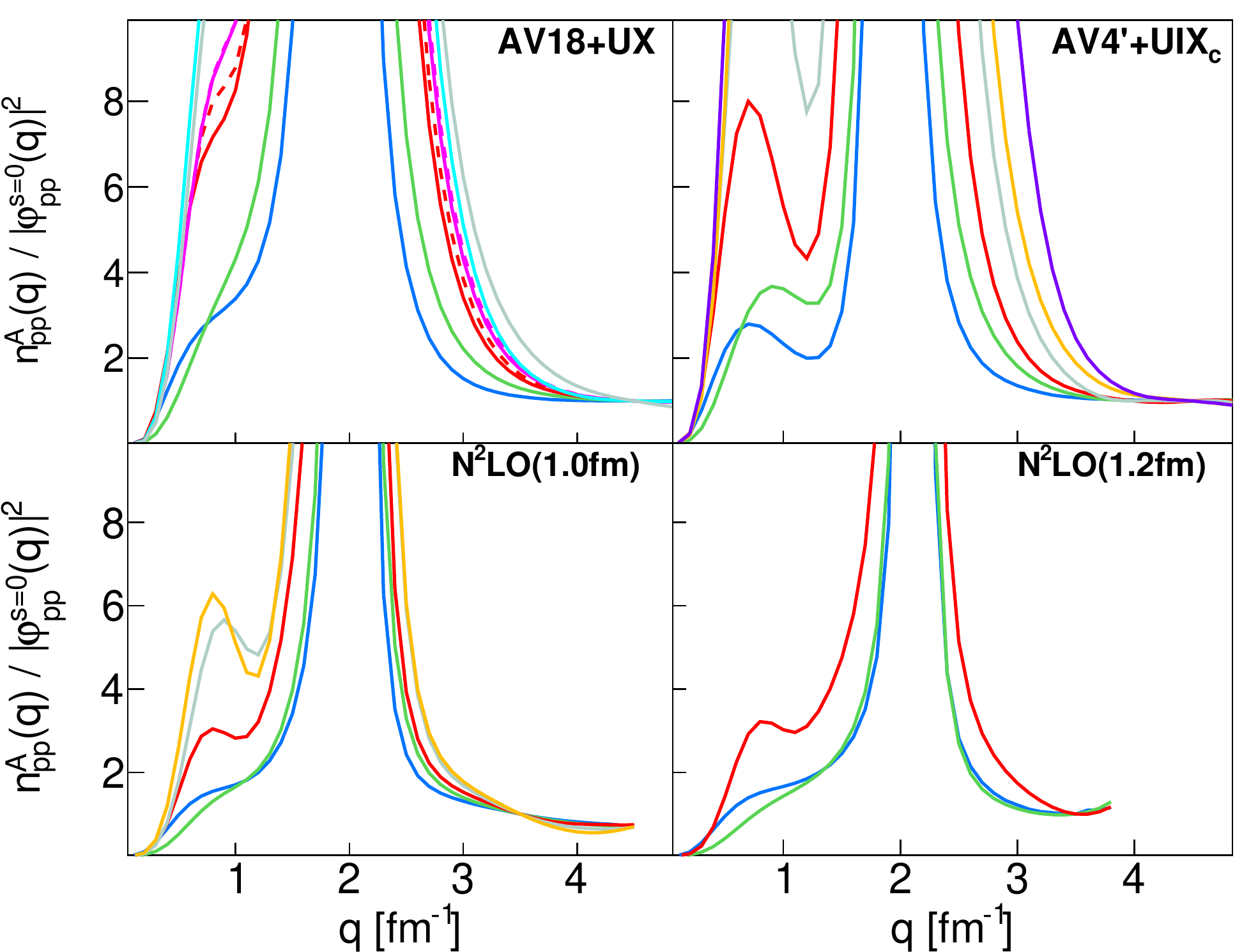} 
\caption{Same as \cref{Fig:DensityScaling_r} but for the two-nucleon momentum-space distribution ratios, $n_A^{NN,\alpha}(q) / |\varphi_{NN}^{\alpha}(q)|^2$, for $pn$ (left) and $pp$ (right) pairs, calculated using different $NN$+$3N$ potentials for different nuclei. See text for details. }
\label{Fig:DensityScaling_k}
\end{figure*}

\section{Two-Body Factorization at Short-Distance \& High-Momentum}

We first study the validity of \cref{Eq1} for describing the two-nucleon distributions calculated using each $NN$+$3N$ potential. \cref{Fig:DensityScaling_r} shows the spatial density $\rho_A^{NN,\alpha}(r)$ for $pn$ and $pp$ pairs for all four interactions and different nuclei, together with the two-body universal functions (see Supplementary Materials for $nn$ distributions). 

While QMC results for each interaction exhibit a specific short-distance behavior, this behavior is the same for all nuclei, and it is consistent with that of the two-body universal functions. This validates the factorization of \cref{Eq1}. The scale factors used to make QMC results of different nuclei agree up to $\sim 1\,\rm fm$ correspond to the $C_A^{NN, \alpha}$ coefficients. We also note that excluding 3N forces in the QMC calculation does not impact significantly the observed scaling at short-distance.

The equivalent study of two-nucleon momentum distributions is more delicate. SRC pairs have high relative momentum, smaller c.m.\ momentum, and small separation. They dominate the nucleon momentum distribution starting at $\sim k_F$~\cite{Hen:2016kwk, Atti:2015eda}. The presently calculated two-nucleon relative momentum distributions however, include contributions from all possible pairs of nucleons, not just short-distance pairs. This includes uncorrelated nucleon pairs, with both high relative and high c.m. momenta. To exclude these, one must consider scaling at much higher relative momenta, where the effect of such uncorrelated pairs is suppressed, see discussion in Ref.~\cite{Weiss:2016obx} and online Supplementary Materials for details.

Since the two-nucleon momentum distributions $n_A^{NN,\alpha}(q)$ decay exponentially, \cref{Fig:DensityScaling_k} shows the ratio $n_A^{NN,\alpha}(q) / |\varphi_{NN}^{\alpha}(q)|^2$, scaled to a value of one at high momenta ($4.5\,\rm fm^{-1}$ in the case of phenomenological potentials, and $3.5\,\rm fm^{-1}$ in the case of chiral interactions). The N$^2$LO $1.0\,\rm fm$ and $1.2\,\rm fm$ distributions are only shown up to $4.4$ and $3.8\,\rm fm^{-1}$ respectively, above which cutoff effects become very large. Here again,  the scale factors used to make the ratio for all nuclei equal one at high-momentum correspond to the $C_A^{NN, \alpha}$ coefficients.

As expected from \cref{Eq1}, the ratios shown in \cref{Fig:DensityScaling_k} scale, i.e., are constant at high-momenta. For $pn$ pairs the scaling is clear and starts at $3.5-4\,\rm fm^{-1}$ for phenomenological potentials, and at $2-2.5\,\rm fm^{-1}$ for chiral interactions. For $pp$ pairs the scaling is less pronounced but still visible, starting at slightly higher momenta than the equivalent $pn$ scaling. As expected, the scaling onset for  two-nucleon momentum distributions is higher than $k_F$ due to contributions from uncorrelated pairs~\cite{Weiss:2016obx}. The scaling of the N$^2$LO calculations starts earlier due to their softer universal functions which suppress the high-momenta uncorrelated pairs (online Supplementary Materials for details).


\section{Scale \& Scheme Independence of Contact Ratios}
The GCF factorization (Eq.~\ref{Eq1}) includes the scale and scheme dependent two-body universal function $\varphi_{NN}^{\alpha}$ and the nuclear contact $C_A^{NN, \alpha}$. The latter encapsulates the many-body dynamics driving the formation of SRCs and needs to be seperated from the two-body part in order to study its scale and scheme dependence. To this end, we examine the ratio of two-nucleon distributions in nucleus $A$ relative to a reference nucleus $A_0$, i.e., $\rho_A^{NN,\alpha}(r) / \rho_{A_0}^{NN,\alpha}(r)$ and $n_A^{NN,\alpha}(q) / n_{A_0}^{NN,\alpha}(q)$. According to \cref{Eq1}, in the scaling region of small-$r$ or large-$q$ such ratios should be independent of $r$ or $q$ and equal $C_A^{NN, \alpha}/C_{A_0}^{NN, \alpha}$. These ratios allow a study of the $A$-dependence of the contact terms, independent from the universal functions~\cite{Weiss:2019}.

\begin{figure}[b]
\includegraphics[width=\columnwidth]{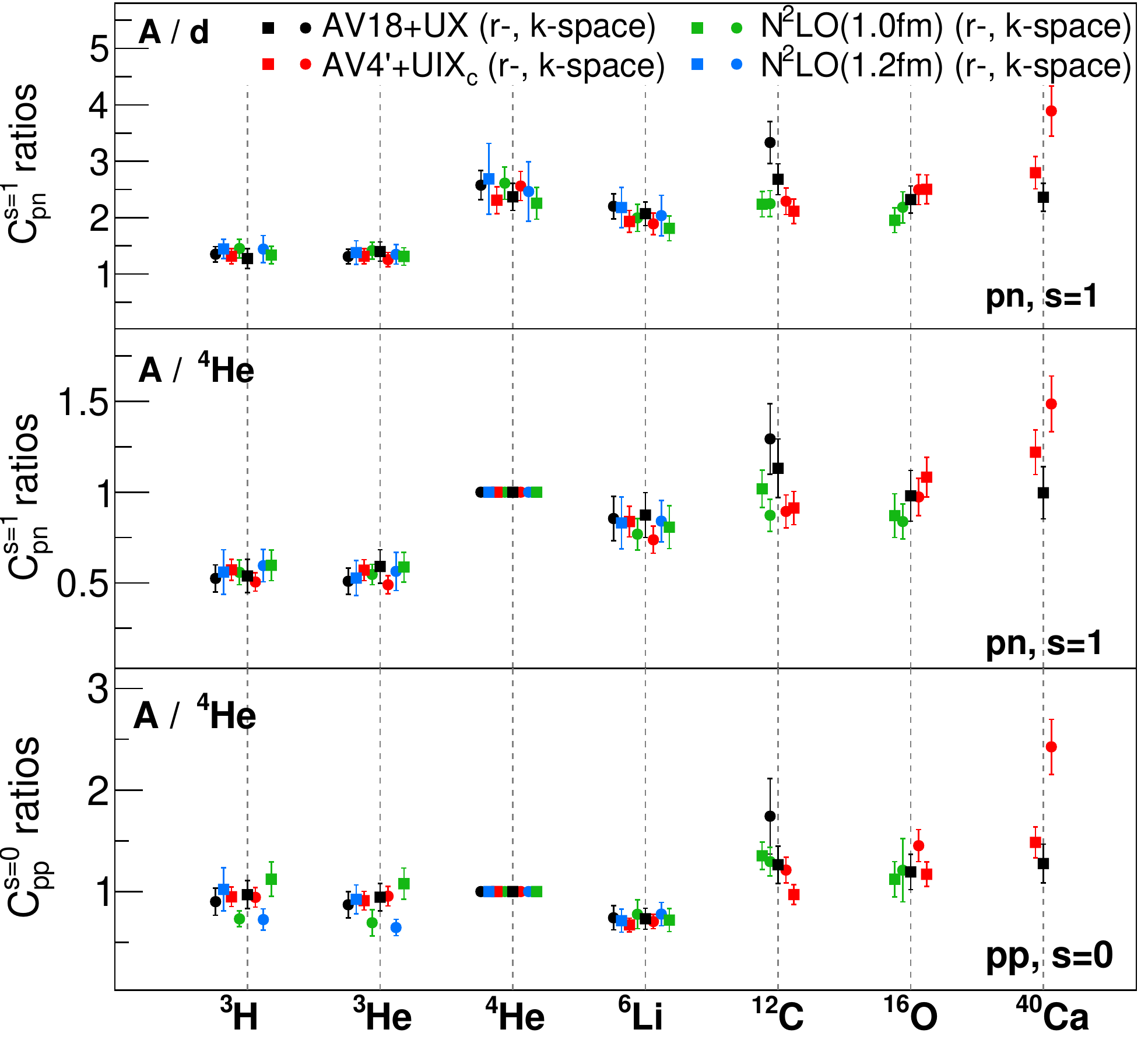}
\caption{Ratios of spin-1 $pn$ contact terms for different nuclei to deuterium (top) or $^4$He (middle), and of spin-0 $pp$ contact terms for different nuclei to $^4$He (bottom). The contact terms ratios were extracted using different $NN$+$3N$ potentials in both coordinate (squares) and momentum (circles) space. The contact values for $^3$H in the spin-0 $pp$ panel corresponds to $C_{nn}^{s=0}$, as there are no pp pairs in this nucleus. See text for details.}
\label{Fig:ContactScaling}
\end{figure}

\cref{Fig:ContactScaling} shows the extracted contact term ratios $C_A^{pn, s=1}/C_{d}^{pn, s=1}$ and $C_A^{NN, \alpha}/C_{^4\mathrm{He}}^{NN, \alpha}$ for all available nuclei and interactions. These ratios were extracted independently for both coordinate and momentum space by fitting the two-nucleon density ratios at short-distance or high-momentum respectively, following Ref.~\cite{Weiss:2016obx}. The uncertainty shown includes contributions from the sensitivity of the extracted contacts to the fit range, and from the difference between the extraction using VMC, DMC, and extrapolated $\rho_A^{NN,\alpha}(r)$ results. We conservatively fixed the latter uncertainty at 10\% (1$\sigma$), and applied it to all extracted contacts. For N$^2$LO potentials, the difference in contacts extracted by using different models of $3N$ forces (namely, $E\tau$ and $E\mathbbm1$, see Ref.~\cite{Lonardoni:2018prc}) was also included in the uncertainty estimate. See online Supplementary Materials for details.

As can be seen, all contact ratios for a given nucleus, including the highly asymmetric \isotope[3]{H} and \isotope[3]{He} nuclei, are consistent within uncertainties, i.e., are largely scale and scheme independent. They are also the same for both short-distance and high-momentum pairs.

This observation has several implications. The fact that models with very different short-range physics, including the tensor-less AV4', all lead to the same contact-term ratios implies that such ratios are insensitive to the nature of the $NN$+$3N$ interaction. 

Conversely, this insensitivity can be exploited in order to estimate properties of heavier nuclei. Calculations of the contact-term ratio $C_{A}^{NN, \alpha}/C_{A_0}^{NN, \alpha}$ (where $A_0 = d$ or \isotope[4]{He}) can be made using ``soft'' interactions that are more amenable to computation. This ratio may then be multiplied by $C_{A_0}^{NN, \alpha}$ for $d$ or \isotope[4]{He} calculated using a harder interaction to effectively obtain $C_{A}^{NN, \alpha}$ for any nucleus for that hard interaction.


On the experimental side, measurements of inclusive $(e,e')$ scattering cross-section ratios at large momentum transfers and high-$x_B$ for a nucleus $A$ relative to deuterium, $a_2(A/d)$, are traditionally interpreted as indicating the relative abundance of SRC pairs in nucleus $A$ relative to deuterium~\cite{frankfurt93,egiyan02,egiyan06,fomin12}. Comparisons of such measurements in a range of symmetric and asymmetric nuclei are claimed to be sensitive to the $NN$+$3N$ interaction~\cite{Fomin:2017ydn}.

Ref.~\cite{Lynn:2019} showed that the relative abundance of short-distance $NN$ pairs in nucleus $A$ relative to deuterium (i.e., $\rho_A(r)/\rho_d(r)$ for $r \rightarrow 0$, where $\rho_A(r)$ includes all $NN$ pairs) is insensitive to the nuclear interaction, and is numerically consistent with the experimental values of $a_2(A/d)$ for all nuclei considered. This raised doubts about the sensitivity of $a_2(A/d)$ measurements to the $NN$+$3N$ interaction. However, the assumed connection between the  $a_2(A/d)$ data analyzed in momentum space and the calculated pair-distance distributions needed to be justified.

We bolster and extend these observations by showing that the calculated contact ratios are independent of the $NN$+$3N$ interaction in both coordinate and momentum space and for each pair quantum state separately. This is consistent with previous calculations that found the relative abundance of SRC pairs to be a mean-field property of the nuclear medium, with only their specific properties (isospin structure, relative momentum distribution, etc.) being determined by the short distance part of the $NN$ interaction~\cite{vanhalst12,Colle:2013nna,colle15,ryckebusch15,Ryckebusch:2018rct}.  Thus our results raise even more doubts about the connection between the $a_2(A/d)$ measurements and the $NN$+$3N$ interaction.  

\begin{figure}[t]
\includegraphics[width=0.45\textwidth]{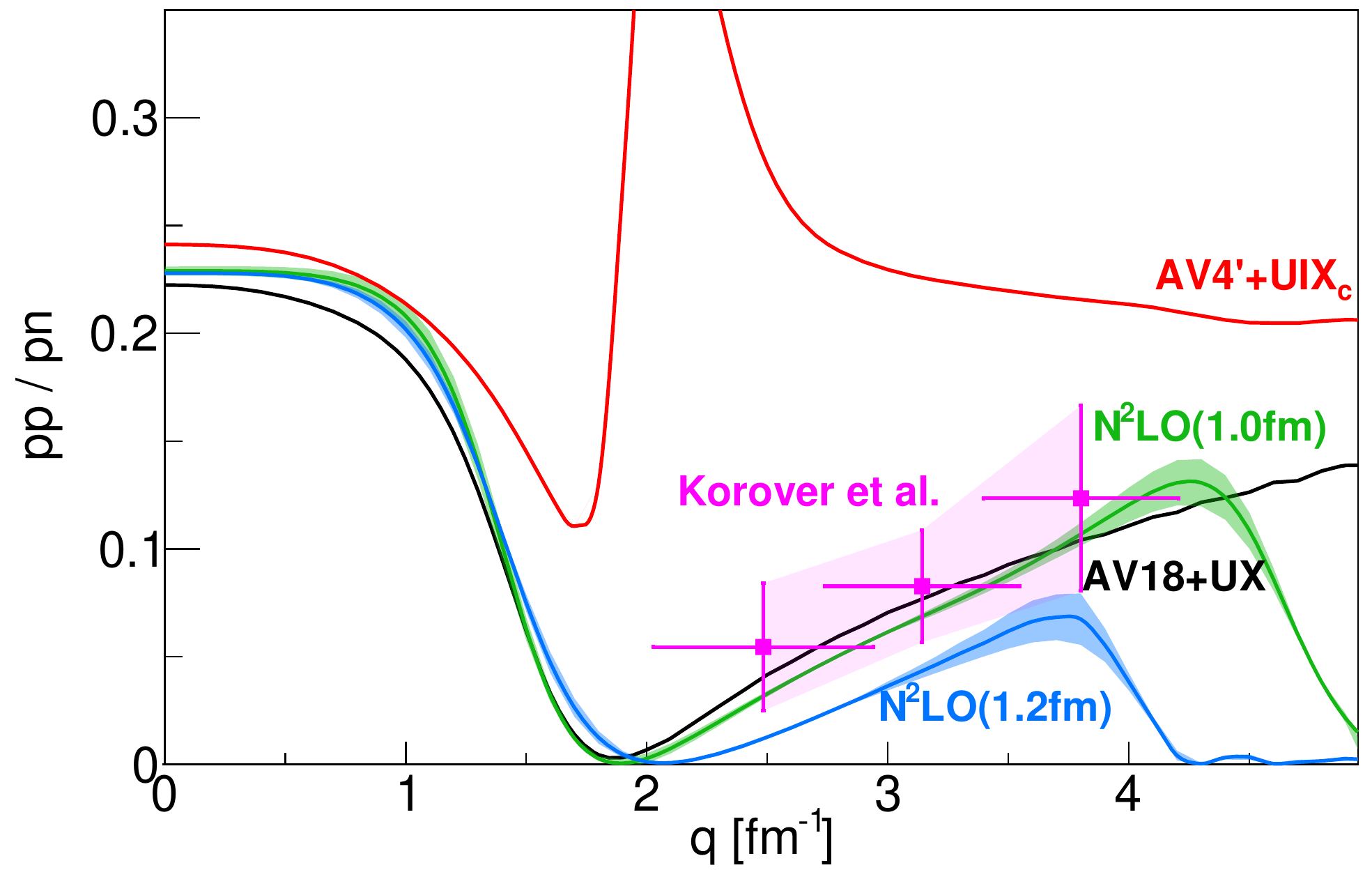}
\caption{Ratio of $pp$-to-$pn$ back-to-back pairs in \isotope[4]{He} as a function of pair relative momentum $q$, $n_A^{pp}(q,Q=0)/n_A^{pn}(q,Q=0)$, for different $NN$+$3N$ potentials, compared with the experimental extractions of Ref.~\cite{korover14} using $(e,e'pp)$ and $(e,e'pn)$ data. See text for details.}
\label{Fig:He4_pp_np}
\end{figure}

An alternate method for probing SRCs is by measuring exclusive two-nucleon knockout reactions $A(e,e'NN)$~\cite{tang03,piasetzky06,subedi08,baghdasaryan10,korover14,hen14,duer18, Duer:2018sxh,Hen:2016kwk}. \cref{Fig:He4_pp_np} shows the ratio of $pp$ to $pn$ pairs in \isotope[4]{He}, as a function of the pair relative momenta, extracted from $(e,e'pp)$ and $(e,e'pn)$ data~\cite{korover14}. The data are compared with two-nucleon distribution ratios $n_{^4\mathrm{He}}^{pp}(q) / n_{^4\mathrm{He}}^{pn}(q)$ calculated at zero c.m. pair momentum ($Q=0$) using different interaction models. Requiring low or zero c.m. momentum (back-to-back pairs) reduces contributions from uncorrelated pairs~\cite{Weiss:2016obx}, allowing to meaningfully compare SRC calculations and measurements. The $pp$ to $pn$ ratio calculated with the AV4'+UIX$_{\rm c}$ interaction is inconsistent with the other calculations and with the experimental data, due to its lack of a tensor force. Thus, exclusive observables can be sensitive to short-distance properties of the nuclear interaction.

Lastly, Refs.~\cite{Arrington12,Arrington:2019wky} claimed there exists a difference between the scaling of SRC pairs with small separation and high relative momenta, and that of pairs with small separation but any relative momenta. The fact that both coordinate- and momentum-space contacts exhibit the same scaling shows that these speculations are inconsistent with QMC wave functions~\cite{Hen:2019jzn}.


\section{ABSOLUTE CONTACTS}

\begin{figure}[t]
\includegraphics[width=\columnwidth]{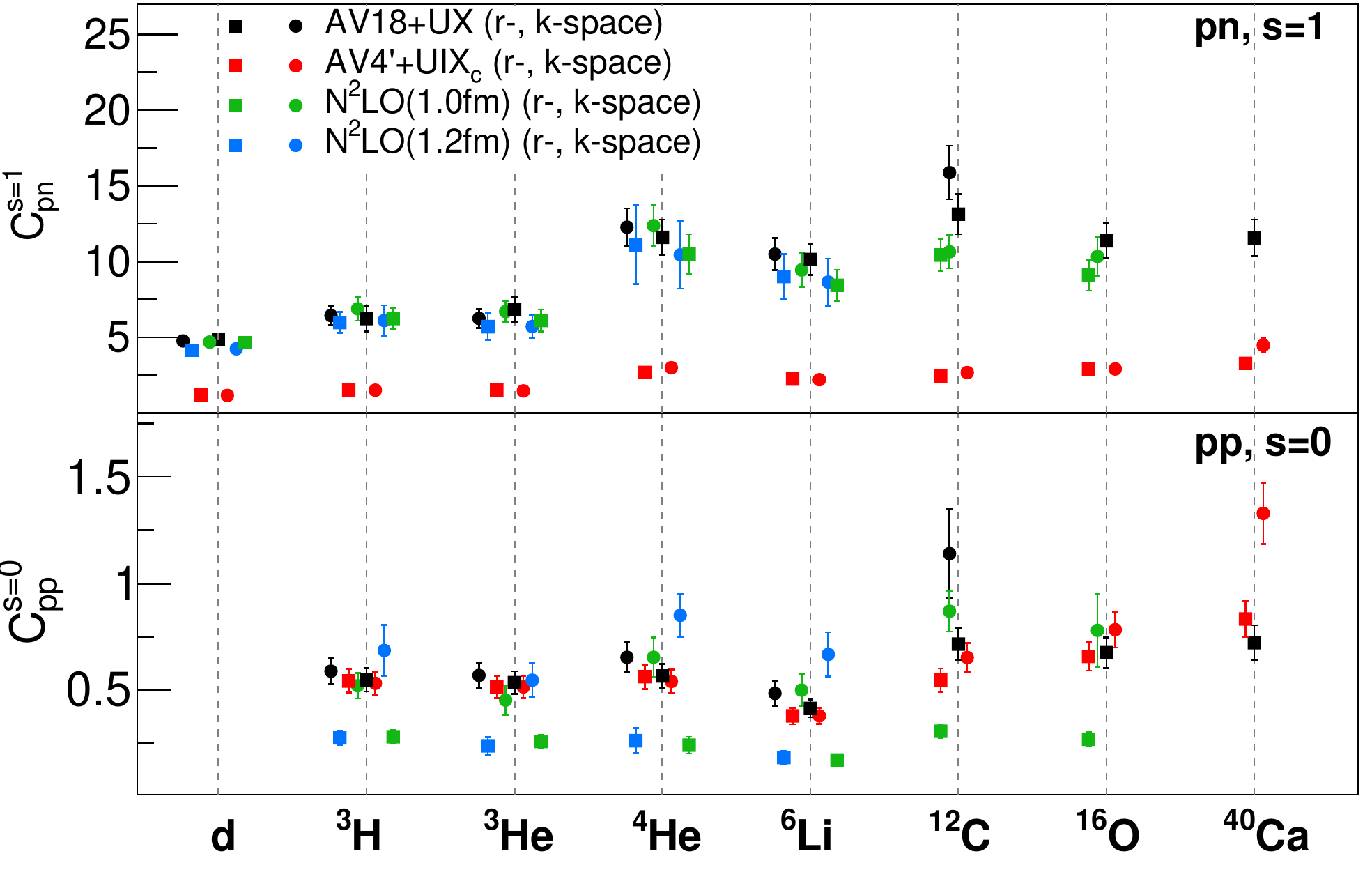}\vspace{0.2cm}
\includegraphics[width=\columnwidth]{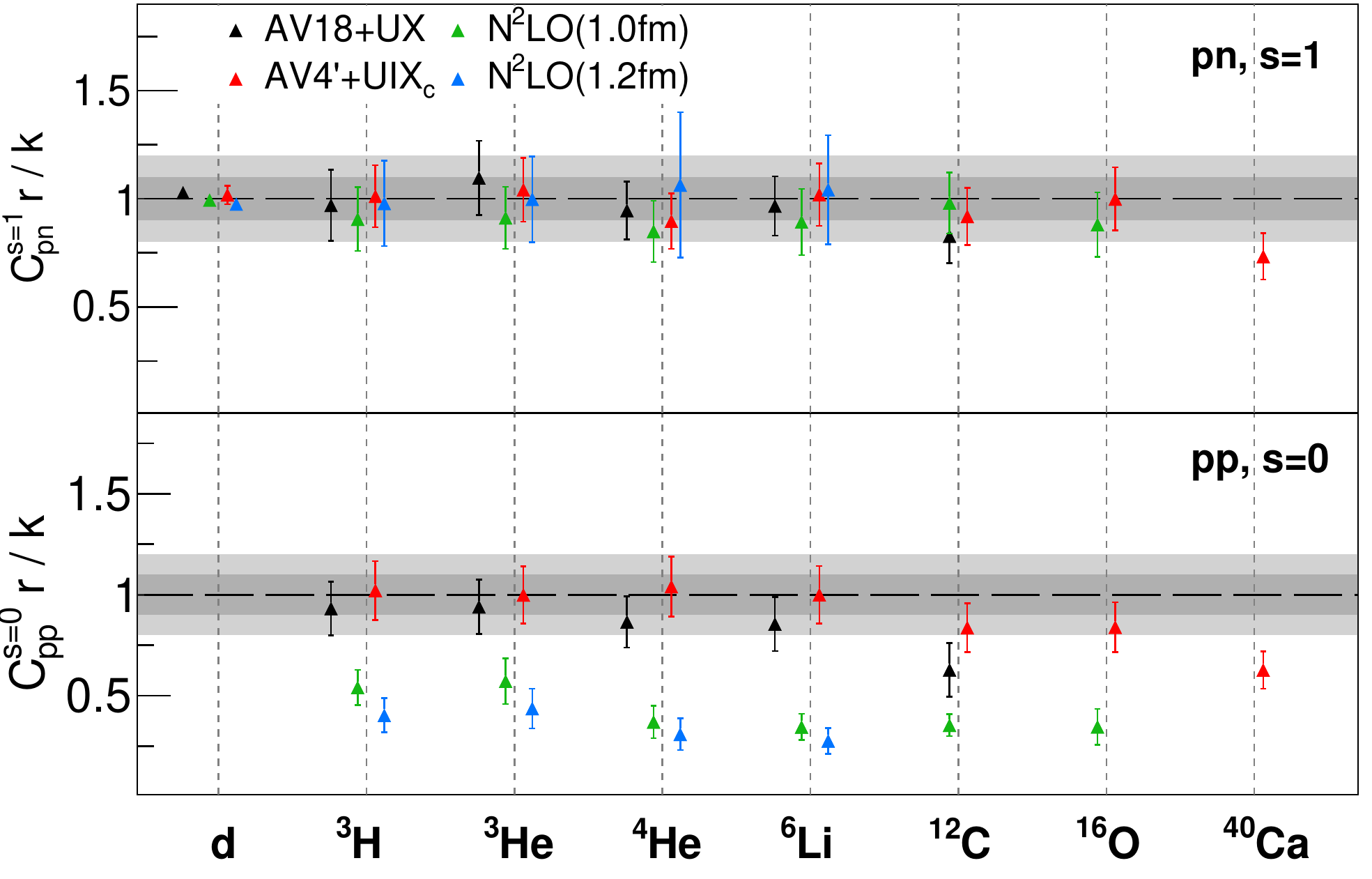}
\caption{Top: The absolute values of the spin-1 $pn$ and spin-0 $pp$ contact terms for different nuclei extracted from QMC calculations in both coordinate (squares) and momentum (circles) space using different $NN$+$3N$ potentials. Bottom: Ratios of coordinate-space to momentum-space contact terms for different nuclei and interaction models. The contact values for $^3$H in the spin-0 $pp$ panel corresponds to $C_{nn}^{s=0}$, as there are no pp pairs in this nucleus. See text for details.}
\label{Fig:Contacts_KspaceNorm}
\end{figure}

Having observed that the {\it ratio} of contact terms for heavier to light nuclei is both scale and scheme independent, and is the same for both small-separation and high-momentum pairs, we now examine the {\it individual} contacts.

\cref{Fig:Contacts_KspaceNorm} (top panel) shows the contacts extracted by fitting \cref{Eq1} to the individual two-nucleon QMC densities for different nuclei in either coordinate or momentum space. In contrast to the contact ratios, here the universal functions do not cancel, so we fixed the normalization of $|\varphi_{NN}^{\alpha}(q)|^2$ so that its integral above $1.3\,\rm fm^{-1} (\approx k_F)$ equals one. This defines the normalization of $|\varphi_{NN}^{\alpha}(r)|^2$ via a Fourier transform, leaving $C_A^{NN, \alpha}$ as the only free fit parameter (see online Supplementary Materials for details).

As can be seen, the extracted contacts are scale and scheme independent (i.e., the same) for all interactions in all channels, except for the AV4'+UIX$_{\rm c}$ spin-1 $pn$ contacts and N$^2$LO spin-0 $pp$ $r$-space contacts (for both $1.0$ and $1.2\,\rm fm$ cutoffs). We expected that the AV4'+UIX$_{\rm c}$ spin-1 $pn$ contacts would be smaller than the AV18+UX and N$^2$LO contacts because AV4' lacks a tensor interaction. The reason for the behavior of the N$^2$LO $r$-space spin-0 $pp$ contacts is less well understood.

To further explore this, \cref{Fig:Contacts_KspaceNorm} (bottom panel) shows the ratio of the $r$- to $k$-space contacts for each nucleus, channel, and interaction. The $r$- and $k$-space contacts are consistent with each other for all cases, except for the N$^2$LO spin-0 $pp$ channel. This discrepancy is not expected in the GCF and will be the focus of future studies. 

Since the contact ratios of nucleus $A$ to \isotope[4]{He} are the same in $r$- and $k$-space for all channels and interactions, including N$^2$LO $pp$ pairs, {\it the many-body nuclear dynamics of the contact terms are the same for both coordinate and momentum space}. Therefore, the N$^2$LO $pp$ issue is possibly due to its (scale and scheme dependent) two-body universal functions which have a non-trivial relation between short-distance and high-momentum pairs. This might be further studied using Wigner distributions that can provide insight to $r$-$k$ correlations in the two-body system~\cite{Neff:2016ajx}.


\section{Summary and Conclusions}
The scale and scheme dependence of SRCs was studied via a GCF analysis of ab-initio QMC calculations of different nuclei using various two- and three-body interaction models. QMC-calculated two-nucleon distributions exhibit significant scale and scheme {\it dependence} at small separation and/or high momentum, as expected from the GCF factorization to a two-body universal function times a contact term.

The extracted absolute contacts are  the same for both short-distance and high-momentum pairs for all interactions except for N$^2$LO in the $pp$ channel. In the latter case, differences are probably due to the relation between the two-body universal functions in $r$- and $k$-space rather than the different many-body dynamics of the contacts, which are both scale and scheme independent and show full $r$- and $k$-space equivalence.

The extracted many-body ratios of contact coefficients for all nuclei relative to $d$ or \isotope{4}{He} are scale and scheme {\it independent} and are also the same for both short-distance and high-momentum pairs. Thus, the formation of {\it short-range} pairs in the nucleus, which determines their abundances, is a {\it long-range} (mean-field) phenomena. Measurements of pair abundances relative to light nuclei are therefore insensitive to short-distance physics.  However, short distance physics can be studied using exclusive two-nucleon knockout measurements.


\begin{acknowledgments}
We thank J.~E.~Lynn for providing some of the deuteron momentum distributions. We are also thankful to J.~Carlson, C.~Ciofi degli Atti, W.~Cosyn, S.~Gandolfi, A.~Lovato, G.A.~Miller, J.~Ryckebusch, M.~Sargsian, and M.~Strikman for many insightful discussions.
This work was supported by the U.S. Department of Energy, Office of Science, Office of Nuclear Physics under Award Numbers DE-FG02-94ER40818, DE-FG02-96ER-40960, and DE-AC05-06OR23177 under which Jefferson Science Associates operates the Thomas Jefferson National Accelerator Facility, the Pazy foundation, and the Israeli Science Foundation (Israel) under Grants Nos. 136/12 and 1334/16.
The work of D.L. is supported by the U.S. Department of Energy, Office of Science, Office of Nuclear Physics, under Contract No. DE-SC0013617, and by the NUCLEI SciDAC program.
The work of R.W. is supported by the Clore Foundation.
The work of R.B.W. is supported by the U.S. Department of Energy, Office of Science, Office of Nuclear Physics, under Contract No. DE-AC02-06CH11357, the NUCLEI SciDAC and INCITE programs, and the Argonne Laboratory Computing Resource Center. 
Computational resources have been provided by Los Alamos Open Supercomputing via the Institutional Computing (IC) program, 
by the Argonne Leadership Computing Facility at Argonne National Laboratory, which is supported by the U.S. Department of Energy, Office of Science, under Contract No. DE-AC02- 06CH11357,
and by the National Energy Research Scientific Computing Center (NERSC), which is supported by the U.S. Department of Energy, Office of Science, under Contract No. DE-AC02-05CH11231.
\end{acknowledgments}

\bibliographystyle{apsrev4-1}
\bibliography{references}

\begin{thebibliography}{64}%
\makeatletter
\providecommand \@ifxundefined [1]{%
 \@ifx{#1\undefined}
}%
\providecommand \@ifnum [1]{%
 \ifnum #1\expandafter \@firstoftwo
 \else \expandafter \@secondoftwo
 \fi
}%
\providecommand \@ifx [1]{%
 \ifx #1\expandafter \@firstoftwo
 \else \expandafter \@secondoftwo
 \fi
}%
\providecommand \natexlab [1]{#1}%
\providecommand \enquote  [1]{``#1''}%
\providecommand \bibnamefont  [1]{#1}%
\providecommand \bibfnamefont [1]{#1}%
\providecommand \citenamefont [1]{#1}%
\providecommand \href@noop [0]{\@secondoftwo}%
\providecommand \href [0]{\begingroup \@sanitize@url \@href}%
\providecommand \@href[1]{\@@startlink{#1}\@@href}%
\providecommand \@@href[1]{\endgroup#1\@@endlink}%
\providecommand \@sanitize@url [0]{\catcode `\\12\catcode `\$12\catcode
  `\&12\catcode `\#12\catcode `\^12\catcode `\_12\catcode `\%12\relax}%
\providecommand \@@startlink[1]{}%
\providecommand \@@endlink[0]{}%
\providecommand \url  [0]{\begingroup\@sanitize@url \@url }%
\providecommand \@url [1]{\endgroup\@href {#1}{\urlprefix }}%
\providecommand \urlprefix  [0]{URL }%
\providecommand \Eprint [0]{\href }%
\providecommand \doibase [0]{http://dx.doi.org/}%
\providecommand \selectlanguage [0]{\@gobble}%
\providecommand \bibinfo  [0]{\@secondoftwo}%
\providecommand \bibfield  [0]{\@secondoftwo}%
\providecommand \translation [1]{[#1]}%
\providecommand \BibitemOpen [0]{}%
\providecommand \bibitemStop [0]{}%
\providecommand \bibitemNoStop [0]{.\EOS\space}%
\providecommand \EOS [0]{\spacefactor3000\relax}%
\providecommand \BibitemShut  [1]{\csname bibitem#1\endcsname}%
\let\auto@bib@innerbib\@empty
\bibitem [{\citenamefont {Hen}\ \emph {et~al.}(2017)\citenamefont {Hen},
  \citenamefont {Miller}, \citenamefont {Piasetzky},\ and\ \citenamefont
  {Weinstein}}]{Hen:2016kwk}%
  \BibitemOpen
  \bibfield  {author} {\bibinfo {author} {\bibfnamefont {O.}~\bibnamefont
  {Hen}}, \bibinfo {author} {\bibfnamefont {G.~A.}\ \bibnamefont {Miller}},
  \bibinfo {author} {\bibfnamefont {E.}~\bibnamefont {Piasetzky}}, \ and\
  \bibinfo {author} {\bibfnamefont {L.~B.}\ \bibnamefont {Weinstein}},\ }\href
  {\doibase 10.1103/RevModPhys.89.045002} {\bibfield  {journal} {\bibinfo
  {journal} {Rev. Mod. Phys.}\ }\textbf {\bibinfo {volume} {89}},\ \bibinfo
  {pages} {045002} (\bibinfo {year} {2017})}\BibitemShut {NoStop}%
\bibitem [{\citenamefont {Ciofi~degli Atti}(2015)}]{Atti:2015eda}%
  \BibitemOpen
  \bibfield  {author} {\bibinfo {author} {\bibfnamefont {C.}~\bibnamefont
  {Ciofi~degli Atti}},\ }\href {\doibase 10.1016/j.physrep.2015.06.002}
  {\bibfield  {journal} {\bibinfo  {journal} {Phys. Rept.}\ }\textbf {\bibinfo
  {volume} {590}},\ \bibinfo {pages} {1} (\bibinfo {year} {2015})}\BibitemShut
  {NoStop}%
\bibitem [{\citenamefont {Tang}\ \emph {et~al.}(2003)\citenamefont {Tang} \emph
  {et~al.}}]{tang03}%
  \BibitemOpen
  \bibfield  {author} {\bibinfo {author} {\bibfnamefont {A.}~\bibnamefont
  {Tang}} \emph {et~al.},\ }\href {\doibase 10.1103/PhysRevLett.90.042301}
  {\bibfield  {journal} {\bibinfo  {journal} {Phys. Rev. Lett.}\ }\textbf
  {\bibinfo {volume} {90}},\ \bibinfo {pages} {042301} (\bibinfo {year}
  {2003})}\BibitemShut {NoStop}%
\bibitem [{\citenamefont {Piasetzky}\ \emph {et~al.}(2006)\citenamefont
  {Piasetzky}, \citenamefont {Sargsian}, \citenamefont {Frankfurt},
  \citenamefont {Strikman},\ and\ \citenamefont {Watson}}]{piasetzky06}%
  \BibitemOpen
  \bibfield  {author} {\bibinfo {author} {\bibfnamefont {E.}~\bibnamefont
  {Piasetzky}}, \bibinfo {author} {\bibfnamefont {M.}~\bibnamefont {Sargsian}},
  \bibinfo {author} {\bibfnamefont {L.}~\bibnamefont {Frankfurt}}, \bibinfo
  {author} {\bibfnamefont {M.}~\bibnamefont {Strikman}}, \ and\ \bibinfo
  {author} {\bibfnamefont {J.~W.}\ \bibnamefont {Watson}},\ }\href {\doibase
  10.1103/PhysRevLett.97.162504} {\bibfield  {journal} {\bibinfo  {journal}
  {Phys. Rev. Lett.}\ }\textbf {\bibinfo {volume} {97}},\ \bibinfo {pages}
  {162504} (\bibinfo {year} {2006})}\BibitemShut {NoStop}%
\bibitem [{\citenamefont {Subedi}\ \emph {et~al.}(2008)\citenamefont {Subedi}
  \emph {et~al.}}]{subedi08}%
  \BibitemOpen
  \bibfield  {author} {\bibinfo {author} {\bibfnamefont {R.}~\bibnamefont
  {Subedi}} \emph {et~al.},\ }\href {\doibase 10.1126/science.1156675}
  {\bibfield  {journal} {\bibinfo  {journal} {Science}\ }\textbf {\bibinfo
  {volume} {320}},\ \bibinfo {pages} {1476} (\bibinfo {year}
  {2008})}\BibitemShut {NoStop}%
\bibitem [{\citenamefont {Baghdasaryan}\ \emph {et~al.}(2010)\citenamefont
  {Baghdasaryan} \emph {et~al.}}]{baghdasaryan10}%
  \BibitemOpen
  \bibfield  {author} {\bibinfo {author} {\bibfnamefont {H.}~\bibnamefont
  {Baghdasaryan}} \emph {et~al.} (\bibinfo {collaboration} {CLAS
  Collaboration}),\ }\href {\doibase 10.1103/PhysRevLett.105.222501} {\bibfield
   {journal} {\bibinfo  {journal} {Phys. Rev. Lett.}\ }\textbf {\bibinfo
  {volume} {105}},\ \bibinfo {pages} {222501} (\bibinfo {year}
  {2010})}\BibitemShut {NoStop}%
\bibitem [{\citenamefont {Korover}\ \emph {et~al.}(2014)\citenamefont {Korover}
  \emph {et~al.}}]{korover14}%
  \BibitemOpen
  \bibfield  {author} {\bibinfo {author} {\bibfnamefont {I.}~\bibnamefont
  {Korover}} \emph {et~al.},\ }\href {\doibase 10.1103/PhysRevLett.113.022501}
  {\bibfield  {journal} {\bibinfo  {journal} {Phys. Rev. Lett.}\ }\textbf
  {\bibinfo {volume} {113}},\ \bibinfo {pages} {022501} (\bibinfo {year}
  {2014})}\BibitemShut {NoStop}%
\bibitem [{\citenamefont {Hen}\ \emph {et~al.}(2014)\citenamefont {Hen} \emph
  {et~al.}}]{hen14}%
  \BibitemOpen
  \bibfield  {author} {\bibinfo {author} {\bibfnamefont {O.}~\bibnamefont
  {Hen}} \emph {et~al.} (\bibinfo {collaboration} {{CLAS Collaboration}}),\
  }\href {\doibase 10.1126/science.1256785} {\bibfield  {journal} {\bibinfo
  {journal} {Science}\ }\textbf {\bibinfo {volume} {346}},\ \bibinfo {pages}
  {614} (\bibinfo {year} {2014})}\BibitemShut {NoStop}%
\bibitem [{\citenamefont {Duer}\ \emph {et~al.}(2018)\citenamefont {Duer} \emph
  {et~al.}}]{duer18}%
  \BibitemOpen
  \bibfield  {author} {\bibinfo {author} {\bibfnamefont {M.}~\bibnamefont
  {Duer}} \emph {et~al.} (\bibinfo {collaboration} {CLAS Collaboration}),\
  }\href {\doibase 10.1038/s41586-018-0400-z} {\bibfield  {journal} {\bibinfo
  {journal} {Nature}\ }\textbf {\bibinfo {volume} {560}},\ \bibinfo {pages}
  {617} (\bibinfo {year} {2018})}\BibitemShut {NoStop}%
\bibitem [{\citenamefont {Duer}\ \emph {et~al.}(2019)\citenamefont {Duer} \emph
  {et~al.}}]{Duer:2018sxh}%
  \BibitemOpen
  \bibfield  {author} {\bibinfo {author} {\bibfnamefont {M.}~\bibnamefont
  {Duer}} \emph {et~al.} (\bibinfo {collaboration} {CLAS Collaboration}),\
  }\href {\doibase 10.1103/PhysRevLett.122.172502} {\bibfield  {journal}
  {\bibinfo  {journal} {Phys. Rev. Lett.}\ }\textbf {\bibinfo {volume} {122}},\
  \bibinfo {pages} {172502} (\bibinfo {year} {2019})}\BibitemShut {NoStop}%
\bibitem [{\citenamefont {Schiavilla}\ \emph {et~al.}(2007)\citenamefont
  {Schiavilla}, \citenamefont {Wiringa}, \citenamefont {Pieper},\ and\
  \citenamefont {Carlson}}]{schiavilla07}%
  \BibitemOpen
  \bibfield  {author} {\bibinfo {author} {\bibfnamefont {R.}~\bibnamefont
  {Schiavilla}}, \bibinfo {author} {\bibfnamefont {R.~B.}\ \bibnamefont
  {Wiringa}}, \bibinfo {author} {\bibfnamefont {S.~C.}\ \bibnamefont {Pieper}},
  \ and\ \bibinfo {author} {\bibfnamefont {J.}~\bibnamefont {Carlson}},\ }\href
  {\doibase 10.1103/PhysRevLett.98.132501} {\bibfield  {journal} {\bibinfo
  {journal} {Phys. Rev. Lett.}\ }\textbf {\bibinfo {volume} {98}},\ \bibinfo
  {pages} {132501} (\bibinfo {year} {2007})}\BibitemShut {NoStop}%
\bibitem [{\citenamefont {Alvioli}\ \emph {et~al.}(2008)\citenamefont
  {Alvioli}, \citenamefont {Ciofi~degli Atti},\ and\ \citenamefont
  {Morita}}]{alvioli08}%
  \BibitemOpen
  \bibfield  {author} {\bibinfo {author} {\bibfnamefont {M.}~\bibnamefont
  {Alvioli}}, \bibinfo {author} {\bibfnamefont {C.}~\bibnamefont {Ciofi~degli
  Atti}}, \ and\ \bibinfo {author} {\bibfnamefont {H.}~\bibnamefont {Morita}},\
  }\href {\doibase 10.1103/PhysRevLett.100.162503} {\bibfield  {journal}
  {\bibinfo  {journal} {Phys. Rev. Lett.}\ }\textbf {\bibinfo {volume} {100}},\
  \bibinfo {pages} {162503} (\bibinfo {year} {2008})}\BibitemShut {NoStop}%
\bibitem [{\citenamefont {Sargsian}\ \emph {et~al.}(2005)\citenamefont
  {Sargsian}, \citenamefont {Abrahamyan}, \citenamefont {Strikman},\ and\
  \citenamefont {Frankfurt}}]{sargsian05}%
  \BibitemOpen
  \bibfield  {author} {\bibinfo {author} {\bibfnamefont {M.~M.}\ \bibnamefont
  {Sargsian}}, \bibinfo {author} {\bibfnamefont {T.~V.}\ \bibnamefont
  {Abrahamyan}}, \bibinfo {author} {\bibfnamefont {M.~I.}\ \bibnamefont
  {Strikman}}, \ and\ \bibinfo {author} {\bibfnamefont {L.~L.}\ \bibnamefont
  {Frankfurt}},\ }\href {\doibase 10.1103/PhysRevC.71.044615} {\bibfield
  {journal} {\bibinfo  {journal} {Phys. Rev. C}\ }\textbf {\bibinfo {volume}
  {71}},\ \bibinfo {pages} {044615} (\bibinfo {year} {2005})}\BibitemShut
  {NoStop}%
\bibitem [{\citenamefont {Weinstein}\ \emph {et~al.}(2011)\citenamefont
  {Weinstein}, \citenamefont {Piasetzky}, \citenamefont {Higinbotham},
  \citenamefont {Gomez}, \citenamefont {Hen},\ and\ \citenamefont
  {Shneor}}]{weinstein11}%
  \BibitemOpen
  \bibfield  {author} {\bibinfo {author} {\bibfnamefont {L.~B.}\ \bibnamefont
  {Weinstein}}, \bibinfo {author} {\bibfnamefont {E.}~\bibnamefont
  {Piasetzky}}, \bibinfo {author} {\bibfnamefont {D.~W.}\ \bibnamefont
  {Higinbotham}}, \bibinfo {author} {\bibfnamefont {J.}~\bibnamefont {Gomez}},
  \bibinfo {author} {\bibfnamefont {O.}~\bibnamefont {Hen}}, \ and\ \bibinfo
  {author} {\bibfnamefont {R.}~\bibnamefont {Shneor}},\ }\href {\doibase
  10.1103/PhysRevLett.106.052301} {\bibfield  {journal} {\bibinfo  {journal}
  {Phys. Rev. Lett.}\ }\textbf {\bibinfo {volume} {106}},\ \bibinfo {pages}
  {052301} (\bibinfo {year} {2011})}\BibitemShut {NoStop}%
\bibitem [{\citenamefont {Hen}\ \emph {et~al.}(2012)\citenamefont {Hen},
  \citenamefont {Piasetzky},\ and\ \citenamefont {Weinstein}}]{Hen12}%
  \BibitemOpen
  \bibfield  {author} {\bibinfo {author} {\bibfnamefont {O.}~\bibnamefont
  {Hen}}, \bibinfo {author} {\bibfnamefont {E.}~\bibnamefont {Piasetzky}}, \
  and\ \bibinfo {author} {\bibfnamefont {L.~B.}\ \bibnamefont {Weinstein}},\
  }\href {\doibase 10.1103/PhysRevC.85.047301} {\bibfield  {journal} {\bibinfo
  {journal} {Phys. Rev. C}\ }\textbf {\bibinfo {volume} {85}},\ \bibinfo
  {pages} {047301} (\bibinfo {year} {2012})}\BibitemShut {NoStop}%
\bibitem [{\citenamefont {Hen}\ \emph {et~al.}(2013)\citenamefont {Hen},
  \citenamefont {Higinbotham}, \citenamefont {Miller}, \citenamefont
  {Piasetzky},\ and\ \citenamefont {Weinstein}}]{Hen:2013oha}%
  \BibitemOpen
  \bibfield  {author} {\bibinfo {author} {\bibfnamefont {O.}~\bibnamefont
  {Hen}}, \bibinfo {author} {\bibfnamefont {D.~W.}\ \bibnamefont
  {Higinbotham}}, \bibinfo {author} {\bibfnamefont {G.~A.}\ \bibnamefont
  {Miller}}, \bibinfo {author} {\bibfnamefont {E.}~\bibnamefont {Piasetzky}}, \
  and\ \bibinfo {author} {\bibfnamefont {L.~B.}\ \bibnamefont {Weinstein}},\
  }\href {\doibase 10.1142/S0218301313300178} {\bibfield  {journal} {\bibinfo
  {journal} {Int. J. Mod. Phys. E}\ }\textbf {\bibinfo {volume} {22}},\
  \bibinfo {pages} {1330017} (\bibinfo {year} {2013})}\BibitemShut {NoStop}%
\bibitem [{\citenamefont {Schmookler}\ \emph {et~al.}(2019)\citenamefont
  {Schmookler} \emph {et~al.}}]{Schmookler:2019nvf}%
  \BibitemOpen
  \bibfield  {author} {\bibinfo {author} {\bibfnamefont {B.}~\bibnamefont
  {Schmookler}} \emph {et~al.} (\bibinfo {collaboration} {CLAS}),\ }\href
  {\doibase 10.1038/s41586-019-0925-9} {\bibfield  {journal} {\bibinfo
  {journal} {Nature}\ }\textbf {\bibinfo {volume} {566}},\ \bibinfo {pages}
  {354} (\bibinfo {year} {2019})}\BibitemShut {NoStop}%
\bibitem [{\citenamefont {Kortelainen}\ and\ \citenamefont
  {Suhonen}(2007{\natexlab{a}})}]{Kortelainen:2007rh}%
  \BibitemOpen
  \bibfield  {author} {\bibinfo {author} {\bibfnamefont {M.}~\bibnamefont
  {Kortelainen}}\ and\ \bibinfo {author} {\bibfnamefont {J.}~\bibnamefont
  {Suhonen}},\ }\href {\doibase 10.1103/PhysRevC.75.051303} {\bibfield
  {journal} {\bibinfo  {journal} {Phys. Rev. C}\ }\textbf {\bibinfo {volume}
  {75}},\ \bibinfo {pages} {051303} (\bibinfo {year}
  {2007}{\natexlab{a}})}\BibitemShut {NoStop}%
\bibitem [{\citenamefont {Kortelainen}\ and\ \citenamefont
  {Suhonen}(2007{\natexlab{b}})}]{Kortelainen:2007mn}%
  \BibitemOpen
  \bibfield  {author} {\bibinfo {author} {\bibfnamefont {M.}~\bibnamefont
  {Kortelainen}}\ and\ \bibinfo {author} {\bibfnamefont {J.}~\bibnamefont
  {Suhonen}},\ }\href {\doibase 10.1103/PhysRevC.76.024315} {\bibfield
  {journal} {\bibinfo  {journal} {Phys. Rev. C}\ }\textbf {\bibinfo {volume}
  {76}},\ \bibinfo {pages} {024315} (\bibinfo {year}
  {2007}{\natexlab{b}})}\BibitemShut {NoStop}%
\bibitem [{\citenamefont {Menendez}\ \emph {et~al.}(2009)\citenamefont
  {Menendez}, \citenamefont {Poves}, \citenamefont {Caurier},\ and\
  \citenamefont {Nowacki}}]{Menendez:2008jp}%
  \BibitemOpen
  \bibfield  {author} {\bibinfo {author} {\bibfnamefont {J.}~\bibnamefont
  {Menendez}}, \bibinfo {author} {\bibfnamefont {A.}~\bibnamefont {Poves}},
  \bibinfo {author} {\bibfnamefont {E.}~\bibnamefont {Caurier}}, \ and\
  \bibinfo {author} {\bibfnamefont {F.}~\bibnamefont {Nowacki}},\ }\href
  {\doibase 10.1016/j.nuclphysa.2008.12.005} {\bibfield  {journal} {\bibinfo
  {journal} {Nucl. Phys. A}\ }\textbf {\bibinfo {volume} {818}},\ \bibinfo
  {pages} {139} (\bibinfo {year} {2009})}\BibitemShut {NoStop}%
\bibitem [{\citenamefont {Simkovic}\ \emph {et~al.}(2009)\citenamefont
  {Simkovic}, \citenamefont {Faessler}, \citenamefont {Muther}, \citenamefont
  {Rodin},\ and\ \citenamefont {Stauf}}]{Simkovic:2009pp}%
  \BibitemOpen
  \bibfield  {author} {\bibinfo {author} {\bibfnamefont {F.}~\bibnamefont
  {Simkovic}}, \bibinfo {author} {\bibfnamefont {A.}~\bibnamefont {Faessler}},
  \bibinfo {author} {\bibfnamefont {H.}~\bibnamefont {Muther}}, \bibinfo
  {author} {\bibfnamefont {V.}~\bibnamefont {Rodin}}, \ and\ \bibinfo {author}
  {\bibfnamefont {M.}~\bibnamefont {Stauf}},\ }\href {\doibase
  10.1103/PhysRevC.79.055501} {\bibfield  {journal} {\bibinfo  {journal} {Phys.
  Rev. C}\ }\textbf {\bibinfo {volume} {79}},\ \bibinfo {pages} {055501}
  (\bibinfo {year} {2009})}\BibitemShut {NoStop}%
\bibitem [{\citenamefont {Benhar}\ \emph {et~al.}(2014)\citenamefont {Benhar},
  \citenamefont {Biondi},\ and\ \citenamefont {Speranza}}]{Benhar:2014cka}%
  \BibitemOpen
  \bibfield  {author} {\bibinfo {author} {\bibfnamefont {O.}~\bibnamefont
  {Benhar}}, \bibinfo {author} {\bibfnamefont {R.}~\bibnamefont {Biondi}}, \
  and\ \bibinfo {author} {\bibfnamefont {E.}~\bibnamefont {Speranza}},\ }\href
  {\doibase 10.1103/PhysRevC.90.065504} {\bibfield  {journal} {\bibinfo
  {journal} {Phys. Rev. C}\ }\textbf {\bibinfo {volume} {90}},\ \bibinfo
  {pages} {065504} (\bibinfo {year} {2014})}\BibitemShut {NoStop}%
\bibitem [{\citenamefont {Cruz-Torres}\ \emph {et~al.}(2018)\citenamefont
  {Cruz-Torres}, \citenamefont {Schmidt}, \citenamefont {Miller}, \citenamefont
  {Weinstein}, \citenamefont {Barnea}, \citenamefont {Weiss}, \citenamefont
  {Piasetzky},\ and\ \citenamefont {Hen}}]{Cruz-Torres:2017sjy}%
  \BibitemOpen
  \bibfield  {author} {\bibinfo {author} {\bibfnamefont {R.}~\bibnamefont
  {Cruz-Torres}}, \bibinfo {author} {\bibfnamefont {A.}~\bibnamefont
  {Schmidt}}, \bibinfo {author} {\bibfnamefont {G.~A.}\ \bibnamefont {Miller}},
  \bibinfo {author} {\bibfnamefont {L.~B.}\ \bibnamefont {Weinstein}}, \bibinfo
  {author} {\bibfnamefont {N.}~\bibnamefont {Barnea}}, \bibinfo {author}
  {\bibfnamefont {R.}~\bibnamefont {Weiss}}, \bibinfo {author} {\bibfnamefont
  {E.}~\bibnamefont {Piasetzky}}, \ and\ \bibinfo {author} {\bibfnamefont
  {O.}~\bibnamefont {Hen}},\ }\href {\doibase 10.1016/j.physletb.2018.07.069}
  {\bibfield  {journal} {\bibinfo  {journal} {Phys. Lett. B}\ }\textbf
  {\bibinfo {volume} {785}},\ \bibinfo {pages} {304} (\bibinfo {year}
  {2018})}\BibitemShut {NoStop}%
\bibitem [{\citenamefont {Wang}\ \emph {et~al.}(2019)\citenamefont {Wang},
  \citenamefont {Hayes}, \citenamefont {Carlson}, \citenamefont {Dong},
  \citenamefont {Mereghetti}, \citenamefont {Pastore},\ and\ \citenamefont
  {Wiringa}}]{Wang:2019hjy}%
  \BibitemOpen
  \bibfield  {author} {\bibinfo {author} {\bibfnamefont {X.~B.}\ \bibnamefont
  {Wang}}, \bibinfo {author} {\bibfnamefont {A.~C.}\ \bibnamefont {Hayes}},
  \bibinfo {author} {\bibfnamefont {J.}~\bibnamefont {Carlson}}, \bibinfo
  {author} {\bibfnamefont {G.~X.}\ \bibnamefont {Dong}}, \bibinfo {author}
  {\bibfnamefont {E.}~\bibnamefont {Mereghetti}}, \bibinfo {author}
  {\bibfnamefont {S.}~\bibnamefont {Pastore}}, \ and\ \bibinfo {author}
  {\bibfnamefont {R.~B.}\ \bibnamefont {Wiringa}},\ }\href@noop {} {\
  (\bibinfo {year} {2019})},\ \Eprint {http://arxiv.org/abs/1906.06662}
  {arXiv:1906.06662 [nucl-th]} \BibitemShut {NoStop}%
\bibitem [{\citenamefont {Miller}\ \emph {et~al.}(2019)\citenamefont {Miller},
  \citenamefont {Beck}, \citenamefont {May-Tal~Beck}, \citenamefont
  {Weinstein}, \citenamefont {Piasetzky},\ and\ \citenamefont
  {Hen}}]{Miller:2018mfb}%
  \BibitemOpen
  \bibfield  {author} {\bibinfo {author} {\bibfnamefont {G.~A.}\ \bibnamefont
  {Miller}}, \bibinfo {author} {\bibfnamefont {A.}~\bibnamefont {Beck}},
  \bibinfo {author} {\bibfnamefont {S.}~\bibnamefont {May-Tal~Beck}}, \bibinfo
  {author} {\bibfnamefont {L.~B.}\ \bibnamefont {Weinstein}}, \bibinfo {author}
  {\bibfnamefont {E.}~\bibnamefont {Piasetzky}}, \ and\ \bibinfo {author}
  {\bibfnamefont {O.}~\bibnamefont {Hen}},\ }\href {\doibase
  https://doi.org/10.1016/j.physletb.2019.05.010} {\bibfield  {journal}
  {\bibinfo  {journal} {Phys. Lett. B}\ }\textbf {\bibinfo {volume} {793}},\
  \bibinfo {pages} {360} (\bibinfo {year} {2019})}\BibitemShut {NoStop}%
\bibitem [{\citenamefont {Li}\ \emph {et~al.}(2018)\citenamefont {Li},
  \citenamefont {Cai}, \citenamefont {Chen},\ and\ \citenamefont
  {Xu}}]{Li:2018lpy}%
  \BibitemOpen
  \bibfield  {author} {\bibinfo {author} {\bibfnamefont {B.-A.}\ \bibnamefont
  {Li}}, \bibinfo {author} {\bibfnamefont {B.-J.}\ \bibnamefont {Cai}},
  \bibinfo {author} {\bibfnamefont {L.-W.}\ \bibnamefont {Chen}}, \ and\
  \bibinfo {author} {\bibfnamefont {J.}~\bibnamefont {Xu}},\ }\href {\doibase
  10.1016/j.ppnp.2018.01.001} {\bibfield  {journal} {\bibinfo  {journal} {Prog.
  Part. Nucl. Phys.}\ }\textbf {\bibinfo {volume} {99}},\ \bibinfo {pages} {29}
  (\bibinfo {year} {2018})}\BibitemShut {NoStop}%
\bibitem [{\citenamefont {Hen}\ \emph {et~al.}(2015)\citenamefont {Hen},
  \citenamefont {Li}, \citenamefont {Guo}, \citenamefont {Weinstein},\ and\
  \citenamefont {Piasetzky}}]{hen15}%
  \BibitemOpen
  \bibfield  {author} {\bibinfo {author} {\bibfnamefont {O.}~\bibnamefont
  {Hen}}, \bibinfo {author} {\bibfnamefont {B.-A.}\ \bibnamefont {Li}},
  \bibinfo {author} {\bibfnamefont {W.-J.}\ \bibnamefont {Guo}}, \bibinfo
  {author} {\bibfnamefont {L.~B.}\ \bibnamefont {Weinstein}}, \ and\ \bibinfo
  {author} {\bibfnamefont {E.}~\bibnamefont {Piasetzky}},\ }\href {\doibase
  10.1103/PhysRevC.91.025803} {\bibfield  {journal} {\bibinfo  {journal} {Phys.
  Rev. C}\ }\textbf {\bibinfo {volume} {91}},\ \bibinfo {pages} {025803}
  (\bibinfo {year} {2015})}\BibitemShut {NoStop}%
\bibitem [{\citenamefont {Frankfurt}\ \emph {et~al.}(2008)\citenamefont
  {Frankfurt}, \citenamefont {Sargsian},\ and\ \citenamefont
  {Strikman}}]{frankfurt08b}%
  \BibitemOpen
  \bibfield  {author} {\bibinfo {author} {\bibfnamefont {L.}~\bibnamefont
  {Frankfurt}}, \bibinfo {author} {\bibfnamefont {M.}~\bibnamefont {Sargsian}},
  \ and\ \bibinfo {author} {\bibfnamefont {M.}~\bibnamefont {Strikman}},\
  }\href {\doibase 10.1142/S0217751X08041207} {\bibfield  {journal} {\bibinfo
  {journal} {Int. J. Mod. Phys. A}\ }\textbf {\bibinfo {volume} {23}},\
  \bibinfo {pages} {2991} (\bibinfo {year} {2008})}\BibitemShut {NoStop}%
\bibitem [{\citenamefont {Wiringa}\ \emph {et~al.}(2014)\citenamefont
  {Wiringa}, \citenamefont {Schiavilla}, \citenamefont {Pieper},\ and\
  \citenamefont {Carlson}}]{Wiringa:2014}%
  \BibitemOpen
  \bibfield  {author} {\bibinfo {author} {\bibfnamefont {R.~B.}\ \bibnamefont
  {Wiringa}}, \bibinfo {author} {\bibfnamefont {R.}~\bibnamefont {Schiavilla}},
  \bibinfo {author} {\bibfnamefont {S.~C.}\ \bibnamefont {Pieper}}, \ and\
  \bibinfo {author} {\bibfnamefont {J.}~\bibnamefont {Carlson}},\ }\href
  {\doibase 10.1103/PhysRevC.89.024305} {\bibfield  {journal} {\bibinfo
  {journal} {Phys. Rev. C}\ }\textbf {\bibinfo {volume} {89}},\ \bibinfo
  {pages} {024305} (\bibinfo {year} {2014})}\BibitemShut {NoStop}%
\bibitem [{\citenamefont {Carlson}\ \emph {et~al.}(2015)\citenamefont
  {Carlson}, \citenamefont {Gandolfi}, \citenamefont {Pederiva}, \citenamefont
  {Pieper}, \citenamefont {Schiavilla}, \citenamefont {Schmidt},\ and\
  \citenamefont {Wiringa}}]{Carlson:2015}%
  \BibitemOpen
  \bibfield  {author} {\bibinfo {author} {\bibfnamefont {J.}~\bibnamefont
  {Carlson}}, \bibinfo {author} {\bibfnamefont {S.}~\bibnamefont {Gandolfi}},
  \bibinfo {author} {\bibfnamefont {F.}~\bibnamefont {Pederiva}}, \bibinfo
  {author} {\bibfnamefont {S.~C.}\ \bibnamefont {Pieper}}, \bibinfo {author}
  {\bibfnamefont {R.}~\bibnamefont {Schiavilla}}, \bibinfo {author}
  {\bibfnamefont {K.~E.}\ \bibnamefont {Schmidt}}, \ and\ \bibinfo {author}
  {\bibfnamefont {R.~B.}\ \bibnamefont {Wiringa}},\ }\href {\doibase
  10.1103/RevModPhys.87.1067} {\bibfield  {journal} {\bibinfo  {journal} {Rev.
  Mod. Phys.}\ }\textbf {\bibinfo {volume} {87}},\ \bibinfo {pages} {1067}
  (\bibinfo {year} {2015})}\BibitemShut {NoStop}%
\bibitem [{\citenamefont {Lonardoni}\ \emph
  {et~al.}(2018{\natexlab{a}})\citenamefont {Lonardoni}, \citenamefont
  {Gandolfi}, \citenamefont {Wang},\ and\ \citenamefont
  {Carlson}}]{Lonardoni:2018nofk}%
  \BibitemOpen
  \bibfield  {author} {\bibinfo {author} {\bibfnamefont {D.}~\bibnamefont
  {Lonardoni}}, \bibinfo {author} {\bibfnamefont {S.}~\bibnamefont {Gandolfi}},
  \bibinfo {author} {\bibfnamefont {X.~B.}\ \bibnamefont {Wang}}, \ and\
  \bibinfo {author} {\bibfnamefont {J.}~\bibnamefont {Carlson}},\ }\href
  {\doibase 10.1103/PhysRevC.98.014322} {\bibfield  {journal} {\bibinfo
  {journal} {Phys. Rev. C}\ }\textbf {\bibinfo {volume} {98}},\ \bibinfo
  {pages} {014322} (\bibinfo {year} {2018}{\natexlab{a}})}\BibitemShut
  {NoStop}%
\bibitem [{\citenamefont {Weiss}\ \emph {et~al.}(2015)\citenamefont {Weiss},
  \citenamefont {Bazak},\ and\ \citenamefont {Barnea}}]{Weiss:2015mba}%
  \BibitemOpen
  \bibfield  {author} {\bibinfo {author} {\bibfnamefont {R.}~\bibnamefont
  {Weiss}}, \bibinfo {author} {\bibfnamefont {B.}~\bibnamefont {Bazak}}, \ and\
  \bibinfo {author} {\bibfnamefont {N.}~\bibnamefont {Barnea}},\ }\href
  {\doibase 10.1103/PhysRevC.92.054311} {\bibfield  {journal} {\bibinfo
  {journal} {Phys. Rev. C}\ }\textbf {\bibinfo {volume} {92}},\ \bibinfo
  {pages} {054311} (\bibinfo {year} {2015})}\BibitemShut {NoStop}%
\bibitem [{\citenamefont {Weiss}\ \emph {et~al.}(2018)\citenamefont {Weiss},
  \citenamefont {Cruz-Torres}, \citenamefont {Barnea}, \citenamefont
  {Piasetzky},\ and\ \citenamefont {Hen}}]{Weiss:2016obx}%
  \BibitemOpen
  \bibfield  {author} {\bibinfo {author} {\bibfnamefont {R.}~\bibnamefont
  {Weiss}}, \bibinfo {author} {\bibfnamefont {R.}~\bibnamefont {Cruz-Torres}},
  \bibinfo {author} {\bibfnamefont {N.}~\bibnamefont {Barnea}}, \bibinfo
  {author} {\bibfnamefont {E.}~\bibnamefont {Piasetzky}}, \ and\ \bibinfo
  {author} {\bibfnamefont {O.}~\bibnamefont {Hen}},\ }\href {\doibase
  https://doi.org/10.1016/j.physletb.2018.01.061} {\bibfield  {journal}
  {\bibinfo  {journal} {Phys. Lett. B}\ }\textbf {\bibinfo {volume} {780}},\
  \bibinfo {pages} {211 } (\bibinfo {year} {2018})}\BibitemShut {NoStop}%
\bibitem [{\citenamefont {Weiss}\ \emph
  {et~al.}(2019{\natexlab{a}})\citenamefont {Weiss}, \citenamefont {Korover},
  \citenamefont {Piasetzky}, \citenamefont {Hen},\ and\ \citenamefont
  {Barnea}}]{Weiss:2018tbu}%
  \BibitemOpen
  \bibfield  {author} {\bibinfo {author} {\bibfnamefont {R.}~\bibnamefont
  {Weiss}}, \bibinfo {author} {\bibfnamefont {I.}~\bibnamefont {Korover}},
  \bibinfo {author} {\bibfnamefont {E.}~\bibnamefont {Piasetzky}}, \bibinfo
  {author} {\bibfnamefont {O.}~\bibnamefont {Hen}}, \ and\ \bibinfo {author}
  {\bibfnamefont {N.}~\bibnamefont {Barnea}},\ }\href {\doibase
  10.1016/j.physletb.2019.02.019} {\bibfield  {journal} {\bibinfo  {journal}
  {Phys. Lett. B}\ }\textbf {\bibinfo {volume} {791}},\ \bibinfo {pages} {242}
  (\bibinfo {year} {2019}{\natexlab{a}})}\BibitemShut {NoStop}%
\bibitem [{\citenamefont {Lynn}\ \emph {et~al.}()\citenamefont {Lynn},
  \citenamefont {Tews}, \citenamefont {Gandolfi},\ and\ \citenamefont
  {Lovato}}]{Lynn:2019}%
  \BibitemOpen
  \bibfield  {author} {\bibinfo {author} {\bibfnamefont {J.~E.}\ \bibnamefont
  {Lynn}}, \bibinfo {author} {\bibfnamefont {I.}~\bibnamefont {Tews}}, \bibinfo
  {author} {\bibfnamefont {S.}~\bibnamefont {Gandolfi}}, \ and\ \bibinfo
  {author} {\bibfnamefont {A.}~\bibnamefont {Lovato}},\ }\href@noop {} {\
  }\Eprint {http://arxiv.org/abs/1901.04868} {arXiv:1901.04868 [nucl-th]}
  \BibitemShut {NoStop}%
\bibitem [{\citenamefont {Lynn}\ \emph {et~al.}(2017)\citenamefont {Lynn},
  \citenamefont {Tews}, \citenamefont {Carlson}, \citenamefont {Gandolfi},
  \citenamefont {Gezerlis}, \citenamefont {Schmidt},\ and\ \citenamefont
  {Schwenk}}]{Lynn:2017}%
  \BibitemOpen
  \bibfield  {author} {\bibinfo {author} {\bibfnamefont {J.~E.}\ \bibnamefont
  {Lynn}}, \bibinfo {author} {\bibfnamefont {I.}~\bibnamefont {Tews}}, \bibinfo
  {author} {\bibfnamefont {J.}~\bibnamefont {Carlson}}, \bibinfo {author}
  {\bibfnamefont {S.}~\bibnamefont {Gandolfi}}, \bibinfo {author}
  {\bibfnamefont {A.}~\bibnamefont {Gezerlis}}, \bibinfo {author}
  {\bibfnamefont {K.~E.}\ \bibnamefont {Schmidt}}, \ and\ \bibinfo {author}
  {\bibfnamefont {A.}~\bibnamefont {Schwenk}},\ }\href {\doibase
  10.1103/PhysRevC.96.054007} {\bibfield  {journal} {\bibinfo  {journal} {Phys.
  Rev. C}\ }\textbf {\bibinfo {volume} {96}},\ \bibinfo {pages} {054007}
  (\bibinfo {year} {2017})}\BibitemShut {NoStop}%
\bibitem [{\citenamefont {Lonardoni}\ \emph {et~al.}(2017)\citenamefont
  {Lonardoni}, \citenamefont {Lovato}, \citenamefont {Pieper},\ and\
  \citenamefont {Wiringa}}]{Lonardoni:2017}%
  \BibitemOpen
  \bibfield  {author} {\bibinfo {author} {\bibfnamefont {D.}~\bibnamefont
  {Lonardoni}}, \bibinfo {author} {\bibfnamefont {A.}~\bibnamefont {Lovato}},
  \bibinfo {author} {\bibfnamefont {S.~C.}\ \bibnamefont {Pieper}}, \ and\
  \bibinfo {author} {\bibfnamefont {R.~B.}\ \bibnamefont {Wiringa}},\ }\href
  {\doibase 10.1103/PhysRevC.96.024326} {\bibfield  {journal} {\bibinfo
  {journal} {Phys. Rev. C}\ }\textbf {\bibinfo {volume} {96}},\ \bibinfo
  {pages} {024326} (\bibinfo {year} {2017})}\BibitemShut {NoStop}%
\bibitem [{\citenamefont {Lonardoni}\ \emph
  {et~al.}(2018{\natexlab{b}})\citenamefont {Lonardoni}, \citenamefont
  {Carlson}, \citenamefont {Gandolfi}, \citenamefont {Lynn}, \citenamefont
  {Schmidt}, \citenamefont {Schwenk},\ and\ \citenamefont
  {Wang}}]{Lonardoni:2018prl}%
  \BibitemOpen
  \bibfield  {author} {\bibinfo {author} {\bibfnamefont {D.}~\bibnamefont
  {Lonardoni}}, \bibinfo {author} {\bibfnamefont {J.}~\bibnamefont {Carlson}},
  \bibinfo {author} {\bibfnamefont {S.}~\bibnamefont {Gandolfi}}, \bibinfo
  {author} {\bibfnamefont {J.~E.}\ \bibnamefont {Lynn}}, \bibinfo {author}
  {\bibfnamefont {K.~E.}\ \bibnamefont {Schmidt}}, \bibinfo {author}
  {\bibfnamefont {A.}~\bibnamefont {Schwenk}}, \ and\ \bibinfo {author}
  {\bibfnamefont {X.~B.}\ \bibnamefont {Wang}},\ }\href {\doibase
  10.1103/PhysRevLett.120.122502} {\bibfield  {journal} {\bibinfo  {journal}
  {Phys. Rev. Lett.}\ }\textbf {\bibinfo {volume} {120}},\ \bibinfo {pages}
  {122502} (\bibinfo {year} {2018}{\natexlab{b}})}\BibitemShut {NoStop}%
\bibitem [{\citenamefont {Lonardoni}\ \emph
  {et~al.}(2018{\natexlab{c}})\citenamefont {Lonardoni}, \citenamefont
  {Gandolfi}, \citenamefont {Lynn}, \citenamefont {Petrie}, \citenamefont
  {Carlson}, \citenamefont {Schmidt},\ and\ \citenamefont
  {Schwenk}}]{Lonardoni:2018prc}%
  \BibitemOpen
  \bibfield  {author} {\bibinfo {author} {\bibfnamefont {D.}~\bibnamefont
  {Lonardoni}}, \bibinfo {author} {\bibfnamefont {S.}~\bibnamefont {Gandolfi}},
  \bibinfo {author} {\bibfnamefont {J.~E.}\ \bibnamefont {Lynn}}, \bibinfo
  {author} {\bibfnamefont {C.}~\bibnamefont {Petrie}}, \bibinfo {author}
  {\bibfnamefont {J.}~\bibnamefont {Carlson}}, \bibinfo {author} {\bibfnamefont
  {K.~E.}\ \bibnamefont {Schmidt}}, \ and\ \bibinfo {author} {\bibfnamefont
  {A.}~\bibnamefont {Schwenk}},\ }\href {\doibase 10.1103/PhysRevC.97.044318}
  {\bibfield  {journal} {\bibinfo  {journal} {Phys. Rev. C}\ }\textbf {\bibinfo
  {volume} {97}},\ \bibinfo {pages} {044318} (\bibinfo {year}
  {2018}{\natexlab{c}})}\BibitemShut {NoStop}%
\bibitem [{\citenamefont {Lynn}\ \emph {et~al.}(2019)\citenamefont {Lynn},
  \citenamefont {Lonardoni}, \citenamefont {Carlson}, \citenamefont {Chen},
  \citenamefont {Detmold}, \citenamefont {Gandolfi},\ and\ \citenamefont
  {Schwenk}}]{Lynn:2019a2}%
  \BibitemOpen
  \bibfield  {author} {\bibinfo {author} {\bibfnamefont {J.~E.}\ \bibnamefont
  {Lynn}}, \bibinfo {author} {\bibfnamefont {D.}~\bibnamefont {Lonardoni}},
  \bibinfo {author} {\bibfnamefont {J.}~\bibnamefont {Carlson}}, \bibinfo
  {author} {\bibfnamefont {J.-W.}\ \bibnamefont {Chen}}, \bibinfo {author}
  {\bibfnamefont {W.}~\bibnamefont {Detmold}}, \bibinfo {author} {\bibfnamefont
  {S.}~\bibnamefont {Gandolfi}}, \ and\ \bibinfo {author} {\bibfnamefont
  {A.}~\bibnamefont {Schwenk}},\ }\href@noop {} {\  (\bibinfo {year} {2019})},\
  \Eprint {http://arxiv.org/abs/1903.12587} {arXiv:1903.12587 [nucl-th]}
  \BibitemShut {NoStop}%
\bibitem [{\citenamefont {Wiringa}()}]{Wiringa:qmc}%
  \BibitemOpen
  \bibfield  {author} {\bibinfo {author} {\bibfnamefont {R.~B.}\ \bibnamefont
  {Wiringa}},\ }\href@noop {} {}\bibinfo {howpublished} {\emph{Quantum Monte
  Carlo results},
  \url{https://www.phy.anl.gov/theory/research/QMCresults.html}},\ \bibinfo
  {note} {{last update: June 20, 2019}}\BibitemShut {NoStop}%
\bibitem [{\citenamefont {Wiringa}\ \emph {et~al.}(1995)\citenamefont
  {Wiringa}, \citenamefont {Stoks},\ and\ \citenamefont
  {Schiavilla}}]{Wiringa:1995}%
  \BibitemOpen
  \bibfield  {author} {\bibinfo {author} {\bibfnamefont {R.~B.}\ \bibnamefont
  {Wiringa}}, \bibinfo {author} {\bibfnamefont {V.~G.~J.}\ \bibnamefont
  {Stoks}}, \ and\ \bibinfo {author} {\bibfnamefont {R.}~\bibnamefont
  {Schiavilla}},\ }\href {\doibase 10.1103/PhysRevC.51.38} {\bibfield
  {journal} {\bibinfo  {journal} {Phys. Rev. C}\ }\textbf {\bibinfo {volume}
  {51}},\ \bibinfo {pages} {38} (\bibinfo {year} {1995})}\BibitemShut {NoStop}%
\bibitem [{\citenamefont {Pudliner}\ \emph {et~al.}(1997)\citenamefont
  {Pudliner}, \citenamefont {Pandharipande}, \citenamefont {Carlson},
  \citenamefont {Pieper},\ and\ \citenamefont {Wiringa}}]{Pudliner:1997}%
  \BibitemOpen
  \bibfield  {author} {\bibinfo {author} {\bibfnamefont {B.~S.}\ \bibnamefont
  {Pudliner}}, \bibinfo {author} {\bibfnamefont {V.~R.}\ \bibnamefont
  {Pandharipande}}, \bibinfo {author} {\bibfnamefont {J.}~\bibnamefont
  {Carlson}}, \bibinfo {author} {\bibfnamefont {S.~C.}\ \bibnamefont {Pieper}},
  \ and\ \bibinfo {author} {\bibfnamefont {R.~B.}\ \bibnamefont {Wiringa}},\
  }\href {\doibase 10.1103/PhysRevC.56.1720} {\bibfield  {journal} {\bibinfo
  {journal} {Phys. Rev. C}\ }\textbf {\bibinfo {volume} {56}},\ \bibinfo
  {pages} {1720} (\bibinfo {year} {1997})}\BibitemShut {NoStop}%
\bibitem [{\citenamefont {Wiringa}\ and\ \citenamefont
  {Pieper}(2002)}]{Wiringa:2002}%
  \BibitemOpen
  \bibfield  {author} {\bibinfo {author} {\bibfnamefont {R.~B.}\ \bibnamefont
  {Wiringa}}\ and\ \bibinfo {author} {\bibfnamefont {S.~C.}\ \bibnamefont
  {Pieper}},\ }\href {\doibase 10.1103/PhysRevLett.89.182501} {\bibfield
  {journal} {\bibinfo  {journal} {Phys. Rev. Lett.}\ }\textbf {\bibinfo
  {volume} {89}},\ \bibinfo {pages} {182501} (\bibinfo {year}
  {2002})}\BibitemShut {NoStop}%
\bibitem [{\citenamefont {Gezerlis}\ \emph {et~al.}(2013)\citenamefont
  {Gezerlis}, \citenamefont {Tews}, \citenamefont {Epelbaum}, \citenamefont
  {Gandolfi}, \citenamefont {Hebeler}, \citenamefont {Nogga},\ and\
  \citenamefont {Schwenk}}]{Gezerlis:2013}%
  \BibitemOpen
  \bibfield  {author} {\bibinfo {author} {\bibfnamefont {A.}~\bibnamefont
  {Gezerlis}}, \bibinfo {author} {\bibfnamefont {I.}~\bibnamefont {Tews}},
  \bibinfo {author} {\bibfnamefont {E.}~\bibnamefont {Epelbaum}}, \bibinfo
  {author} {\bibfnamefont {S.}~\bibnamefont {Gandolfi}}, \bibinfo {author}
  {\bibfnamefont {K.}~\bibnamefont {Hebeler}}, \bibinfo {author} {\bibfnamefont
  {A.}~\bibnamefont {Nogga}}, \ and\ \bibinfo {author} {\bibfnamefont
  {A.}~\bibnamefont {Schwenk}},\ }\href {\doibase
  10.1103/PhysRevLett.111.032501} {\bibfield  {journal} {\bibinfo  {journal}
  {Phys. Rev. Lett.}\ }\textbf {\bibinfo {volume} {111}},\ \bibinfo {pages}
  {032501} (\bibinfo {year} {2013})}\BibitemShut {NoStop}%
\bibitem [{\citenamefont {Gezerlis}\ \emph {et~al.}(2014)\citenamefont
  {Gezerlis}, \citenamefont {Tews}, \citenamefont {Epelbaum}, \citenamefont
  {Freunek}, \citenamefont {Gandolfi}, \citenamefont {Hebeler}, \citenamefont
  {Nogga},\ and\ \citenamefont {Schwenk}}]{Gezerlis:2014}%
  \BibitemOpen
  \bibfield  {author} {\bibinfo {author} {\bibfnamefont {A.}~\bibnamefont
  {Gezerlis}}, \bibinfo {author} {\bibfnamefont {I.}~\bibnamefont {Tews}},
  \bibinfo {author} {\bibfnamefont {E.}~\bibnamefont {Epelbaum}}, \bibinfo
  {author} {\bibfnamefont {M.}~\bibnamefont {Freunek}}, \bibinfo {author}
  {\bibfnamefont {S.}~\bibnamefont {Gandolfi}}, \bibinfo {author}
  {\bibfnamefont {K.}~\bibnamefont {Hebeler}}, \bibinfo {author} {\bibfnamefont
  {A.}~\bibnamefont {Nogga}}, \ and\ \bibinfo {author} {\bibfnamefont
  {A.}~\bibnamefont {Schwenk}},\ }\href {\doibase 10.1103/PhysRevC.90.054323}
  {\bibfield  {journal} {\bibinfo  {journal} {Phys. Rev. C}\ }\textbf {\bibinfo
  {volume} {90}},\ \bibinfo {pages} {054323} (\bibinfo {year}
  {2014})}\BibitemShut {NoStop}%
\bibitem [{\citenamefont {Lynn}\ \emph {et~al.}(2016)\citenamefont {Lynn},
  \citenamefont {Tews}, \citenamefont {Carlson}, \citenamefont {Gandolfi},
  \citenamefont {Gezerlis}, \citenamefont {Schmidt},\ and\ \citenamefont
  {Schwenk}}]{Lynn:2016}%
  \BibitemOpen
  \bibfield  {author} {\bibinfo {author} {\bibfnamefont {J.~E.}\ \bibnamefont
  {Lynn}}, \bibinfo {author} {\bibfnamefont {I.}~\bibnamefont {Tews}}, \bibinfo
  {author} {\bibfnamefont {J.}~\bibnamefont {Carlson}}, \bibinfo {author}
  {\bibfnamefont {S.}~\bibnamefont {Gandolfi}}, \bibinfo {author}
  {\bibfnamefont {A.}~\bibnamefont {Gezerlis}}, \bibinfo {author}
  {\bibfnamefont {K.~E.}\ \bibnamefont {Schmidt}}, \ and\ \bibinfo {author}
  {\bibfnamefont {A.}~\bibnamefont {Schwenk}},\ }\href {\doibase
  10.1103/PhysRevLett.116.062501} {\bibfield  {journal} {\bibinfo  {journal}
  {Phys. Rev. Lett.}\ }\textbf {\bibinfo {volume} {116}},\ \bibinfo {pages}
  {062501} (\bibinfo {year} {2016})}\BibitemShut {NoStop}%
\bibitem [{\citenamefont {Cohen}\ \emph {et~al.}(2018)\citenamefont {Cohen}
  \emph {et~al.}}]{Cohen:2018gzh}%
  \BibitemOpen
  \bibfield  {author} {\bibinfo {author} {\bibfnamefont {E.~O.}\ \bibnamefont
  {Cohen}} \emph {et~al.} (\bibinfo {collaboration} {CLAS Collaboration}),\
  }\href {\doibase 10.1103/PhysRevLett.121.092501} {\bibfield  {journal}
  {\bibinfo  {journal} {Phys. Rev. Lett.}\ }\textbf {\bibinfo {volume} {121}},\
  \bibinfo {pages} {092501} (\bibinfo {year} {2018})}\BibitemShut {NoStop}%
\bibitem [{\citenamefont {Chen}\ \emph {et~al.}(2017)\citenamefont {Chen},
  \citenamefont {Detmold}, \citenamefont {Lynn},\ and\ \citenamefont
  {Schwenk}}]{Chen:2016bde}%
  \BibitemOpen
  \bibfield  {author} {\bibinfo {author} {\bibfnamefont {J.-W.}\ \bibnamefont
  {Chen}}, \bibinfo {author} {\bibfnamefont {W.}~\bibnamefont {Detmold}},
  \bibinfo {author} {\bibfnamefont {J.~E.}\ \bibnamefont {Lynn}}, \ and\
  \bibinfo {author} {\bibfnamefont {A.}~\bibnamefont {Schwenk}},\ }\href
  {\doibase 10.1103/PhysRevLett.119.262502} {\bibfield  {journal} {\bibinfo
  {journal} {Phys. Rev. Lett.}\ }\textbf {\bibinfo {volume} {119}},\ \bibinfo
  {pages} {262502} (\bibinfo {year} {2017})}\BibitemShut {NoStop}%
\bibitem [{\citenamefont {Weiss}\ \emph
  {et~al.}(2019{\natexlab{b}})\citenamefont {Weiss}, \citenamefont {Schmidt},
  \citenamefont {Miller},\ and\ \citenamefont {Barnea}}]{Weiss:2019}%
  \BibitemOpen
  \bibfield  {author} {\bibinfo {author} {\bibfnamefont {R.}~\bibnamefont
  {Weiss}}, \bibinfo {author} {\bibfnamefont {A.}~\bibnamefont {Schmidt}},
  \bibinfo {author} {\bibfnamefont {G.~A.}\ \bibnamefont {Miller}}, \ and\
  \bibinfo {author} {\bibfnamefont {N.}~\bibnamefont {Barnea}},\ }\href
  {\doibase https://doi.org/10.1016/j.physletb.2019.01.053} {\bibfield
  {journal} {\bibinfo  {journal} {Phys. Lett. B}\ }\textbf {\bibinfo {volume}
  {790}},\ \bibinfo {pages} {484} (\bibinfo {year}
  {2019}{\natexlab{b}})}\BibitemShut {NoStop}%
\bibitem [{\citenamefont {Frankfurt}\ \emph {et~al.}(1993)\citenamefont
  {Frankfurt}, \citenamefont {Strikman}, \citenamefont {Day},\ and\
  \citenamefont {Sargsyan}}]{frankfurt93}%
  \BibitemOpen
  \bibfield  {author} {\bibinfo {author} {\bibfnamefont {L.~L.}\ \bibnamefont
  {Frankfurt}}, \bibinfo {author} {\bibfnamefont {M.~I.}\ \bibnamefont
  {Strikman}}, \bibinfo {author} {\bibfnamefont {D.~B.}\ \bibnamefont {Day}}, \
  and\ \bibinfo {author} {\bibfnamefont {M.}~\bibnamefont {Sargsyan}},\ }\href
  {\doibase 10.1103/PhysRevC.48.2451} {\bibfield  {journal} {\bibinfo
  {journal} {Phys. Rev. C}\ }\textbf {\bibinfo {volume} {48}},\ \bibinfo
  {pages} {2451} (\bibinfo {year} {1993})}\BibitemShut {NoStop}%
\bibitem [{\citenamefont {Egiyan}\ \emph {et~al.}(2003)\citenamefont {Egiyan}
  \emph {et~al.}}]{egiyan02}%
  \BibitemOpen
  \bibfield  {author} {\bibinfo {author} {\bibfnamefont {K.~S.}\ \bibnamefont
  {Egiyan}} \emph {et~al.} (\bibinfo {collaboration} {CLAS Collaboration}),\
  }\href {\doibase 10.1103/PhysRevC.68.014313} {\bibfield  {journal} {\bibinfo
  {journal} {Phys. Rev. C}\ }\textbf {\bibinfo {volume} {68}},\ \bibinfo
  {pages} {014313} (\bibinfo {year} {2003})}\BibitemShut {NoStop}%
\bibitem [{\citenamefont {Egiyan}\ \emph {et~al.}(2006)\citenamefont {Egiyan}
  \emph {et~al.}}]{egiyan06}%
  \BibitemOpen
  \bibfield  {author} {\bibinfo {author} {\bibfnamefont {K.~S.}\ \bibnamefont
  {Egiyan}} \emph {et~al.} (\bibinfo {collaboration} {CLAS Collaboration}),\
  }\href {\doibase 10.1103/PhysRevLett.96.082501} {\bibfield  {journal}
  {\bibinfo  {journal} {Phys. Rev. Lett.}\ }\textbf {\bibinfo {volume} {96}},\
  \bibinfo {pages} {082501} (\bibinfo {year} {2006})}\BibitemShut {NoStop}%
\bibitem [{\citenamefont {Fomin}\ \emph {et~al.}(2012)\citenamefont {Fomin}
  \emph {et~al.}}]{fomin12}%
  \BibitemOpen
  \bibfield  {author} {\bibinfo {author} {\bibfnamefont {N.}~\bibnamefont
  {Fomin}} \emph {et~al.},\ }\href {\doibase 10.1103/PhysRevLett.108.092502}
  {\bibfield  {journal} {\bibinfo  {journal} {Phys. Rev. Lett.}\ }\textbf
  {\bibinfo {volume} {108}},\ \bibinfo {pages} {092502} (\bibinfo {year}
  {2012})}\BibitemShut {NoStop}%
\bibitem [{\citenamefont {Fomin}\ \emph {et~al.}(2017)\citenamefont {Fomin},
  \citenamefont {Higinbotham}, \citenamefont {Sargsian},\ and\ \citenamefont
  {Solvignon}}]{Fomin:2017ydn}%
  \BibitemOpen
  \bibfield  {author} {\bibinfo {author} {\bibfnamefont {N.}~\bibnamefont
  {Fomin}}, \bibinfo {author} {\bibfnamefont {D.}~\bibnamefont {Higinbotham}},
  \bibinfo {author} {\bibfnamefont {M.}~\bibnamefont {Sargsian}}, \ and\
  \bibinfo {author} {\bibfnamefont {P.}~\bibnamefont {Solvignon}},\ }\href
  {\doibase 10.1146/annurev-nucl-102115-044939} {\bibfield  {journal} {\bibinfo
   {journal} {Ann. Rev. Nucl. Part. Sci.}\ }\textbf {\bibinfo {volume} {67}},\
  \bibinfo {pages} {129} (\bibinfo {year} {2017})}\BibitemShut {NoStop}%
\bibitem [{\citenamefont {Vanhalst}\ \emph {et~al.}(2012)\citenamefont
  {Vanhalst}, \citenamefont {Ryckebusch},\ and\ \citenamefont
  {Cosyn}}]{vanhalst12}%
  \BibitemOpen
  \bibfield  {author} {\bibinfo {author} {\bibfnamefont {M.}~\bibnamefont
  {Vanhalst}}, \bibinfo {author} {\bibfnamefont {J.}~\bibnamefont
  {Ryckebusch}}, \ and\ \bibinfo {author} {\bibfnamefont {W.}~\bibnamefont
  {Cosyn}},\ }\href {\doibase 10.1103/PhysRevC.86.044619} {\bibfield  {journal}
  {\bibinfo  {journal} {Phys. Rev. C}\ }\textbf {\bibinfo {volume} {86}},\
  \bibinfo {pages} {044619} (\bibinfo {year} {2012})}\BibitemShut {NoStop}%
\bibitem [{\citenamefont {Colle}\ \emph {et~al.}(2014)\citenamefont {Colle},
  \citenamefont {Cosyn}, \citenamefont {Ryckebusch},\ and\ \citenamefont
  {Vanhalst}}]{Colle:2013nna}%
  \BibitemOpen
  \bibfield  {author} {\bibinfo {author} {\bibfnamefont {C.}~\bibnamefont
  {Colle}}, \bibinfo {author} {\bibfnamefont {W.}~\bibnamefont {Cosyn}},
  \bibinfo {author} {\bibfnamefont {J.}~\bibnamefont {Ryckebusch}}, \ and\
  \bibinfo {author} {\bibfnamefont {M.}~\bibnamefont {Vanhalst}},\ }\href
  {\doibase 10.1103/PhysRevC.89.024603} {\bibfield  {journal} {\bibinfo
  {journal} {Phys. Rev. C}\ }\textbf {\bibinfo {volume} {89}},\ \bibinfo
  {pages} {024603} (\bibinfo {year} {2014})}\BibitemShut {NoStop}%
\bibitem [{\citenamefont {Colle}\ \emph {et~al.}(2015)\citenamefont {Colle}
  \emph {et~al.}}]{colle15}%
  \BibitemOpen
  \bibfield  {author} {\bibinfo {author} {\bibfnamefont {C.}~\bibnamefont
  {Colle}} \emph {et~al.},\ }\href {\doibase 10.1103/PhysRevC.92.024604}
  {\bibfield  {journal} {\bibinfo  {journal} {Phys. Rev. C}\ }\textbf {\bibinfo
  {volume} {92}},\ \bibinfo {pages} {024604} (\bibinfo {year}
  {2015})}\BibitemShut {NoStop}%
\bibitem [{\citenamefont {J}\ \emph {et~al.}(2015)\citenamefont {J},
  \citenamefont {Vanhalst},\ and\ \citenamefont {Cosyn}}]{ryckebusch15}%
  \BibitemOpen
  \bibfield  {author} {\bibinfo {author} {\bibfnamefont {R.}~\bibnamefont {J}},
  \bibinfo {author} {\bibfnamefont {M.}~\bibnamefont {Vanhalst}}, \ and\
  \bibinfo {author} {\bibfnamefont {W.}~\bibnamefont {Cosyn}},\ }\href
  {\doibase 10.1088/0954-3899/42/5/055104} {\bibfield  {journal} {\bibinfo
  {journal} {J. Phys. G: Nucl. Part. Phys.}\ }\textbf {\bibinfo {volume}
  {42}},\ \bibinfo {pages} {055104} (\bibinfo {year} {2015})}\BibitemShut
  {NoStop}%
\bibitem [{\citenamefont {Ryckebusch}\ \emph {et~al.}(2019)\citenamefont
  {Ryckebusch}, \citenamefont {Cosyn}, \citenamefont {Stevens}, \citenamefont
  {Casert},\ and\ \citenamefont {Nys}}]{Ryckebusch:2018rct}%
  \BibitemOpen
  \bibfield  {author} {\bibinfo {author} {\bibfnamefont {J.}~\bibnamefont
  {Ryckebusch}}, \bibinfo {author} {\bibfnamefont {W.}~\bibnamefont {Cosyn}},
  \bibinfo {author} {\bibfnamefont {S.}~\bibnamefont {Stevens}}, \bibinfo
  {author} {\bibfnamefont {C.}~\bibnamefont {Casert}}, \ and\ \bibinfo {author}
  {\bibfnamefont {J.}~\bibnamefont {Nys}},\ }\href {\doibase
  10.1016/j.physletb.2019.03.016} {\bibfield  {journal} {\bibinfo  {journal}
  {Phys. Lett. B}\ }\textbf {\bibinfo {volume} {792}},\ \bibinfo {pages} {21}
  (\bibinfo {year} {2019})},\ \Eprint {http://arxiv.org/abs/1808.09859}
  {arXiv:1808.09859 [nucl-th]} \BibitemShut {NoStop}%
\bibitem [{\citenamefont {Arrington}\ \emph {et~al.}(2012)\citenamefont
  {Arrington}, \citenamefont {Daniel}, \citenamefont {Day}, \citenamefont
  {Fomin}, \citenamefont {Gaskell},\ and\ \citenamefont
  {Solvignon}}]{Arrington12}%
  \BibitemOpen
  \bibfield  {author} {\bibinfo {author} {\bibfnamefont {J.}~\bibnamefont
  {Arrington}}, \bibinfo {author} {\bibfnamefont {A.}~\bibnamefont {Daniel}},
  \bibinfo {author} {\bibfnamefont {D.~B.}\ \bibnamefont {Day}}, \bibinfo
  {author} {\bibfnamefont {N.}~\bibnamefont {Fomin}}, \bibinfo {author}
  {\bibfnamefont {D.}~\bibnamefont {Gaskell}}, \ and\ \bibinfo {author}
  {\bibfnamefont {P.}~\bibnamefont {Solvignon}},\ }\href {\doibase
  10.1103/PhysRevC.86.065204} {\bibfield  {journal} {\bibinfo  {journal} {Phys.
  Rev. C}\ }\textbf {\bibinfo {volume} {86}},\ \bibinfo {pages} {065204}
  (\bibinfo {year} {2012})}\BibitemShut {NoStop}%
\bibitem [{\citenamefont {Arrington}\ and\ \citenamefont
  {Fomin}(2019)}]{Arrington:2019wky}%
  \BibitemOpen
  \bibfield  {author} {\bibinfo {author} {\bibfnamefont {J.}~\bibnamefont
  {Arrington}}\ and\ \bibinfo {author} {\bibfnamefont {N.}~\bibnamefont
  {Fomin}},\ }\href@noop {} {\  (\bibinfo {year} {2019})},\ \Eprint
  {http://arxiv.org/abs/1903.12535v1} {arXiv:1903.12535v1 [nucl-ex]}
  \BibitemShut {NoStop}%
\bibitem [{\citenamefont {Hen}\ \emph {et~al.}(2019)\citenamefont {Hen},
  \citenamefont {Hauenstein}, \citenamefont {Higinbotham}, \citenamefont
  {Miller}, \citenamefont {Piasetzky}, \citenamefont {Schmidt}, \citenamefont
  {Segarra}, \citenamefont {Strikman},\ and\ \citenamefont
  {Weinstein}}]{Hen:2019jzn}%
  \BibitemOpen
  \bibfield  {author} {\bibinfo {author} {\bibfnamefont {O.}~\bibnamefont
  {Hen}}, \bibinfo {author} {\bibfnamefont {F.}~\bibnamefont {Hauenstein}},
  \bibinfo {author} {\bibfnamefont {D.~W.}\ \bibnamefont {Higinbotham}},
  \bibinfo {author} {\bibfnamefont {G.~A.}\ \bibnamefont {Miller}}, \bibinfo
  {author} {\bibfnamefont {E.}~\bibnamefont {Piasetzky}}, \bibinfo {author}
  {\bibfnamefont {A.}~\bibnamefont {Schmidt}}, \bibinfo {author} {\bibfnamefont
  {E.~P.}\ \bibnamefont {Segarra}}, \bibinfo {author} {\bibfnamefont
  {M.}~\bibnamefont {Strikman}}, \ and\ \bibinfo {author} {\bibfnamefont
  {L.~B.}\ \bibnamefont {Weinstein}},\ }\href@noop {} {\  (\bibinfo {year}
  {2019})},\ \Eprint {http://arxiv.org/abs/1905.02172} {arXiv:1905.02172
  [nucl-ex]} \BibitemShut {NoStop}%
\bibitem [{\citenamefont {Neff}\ and\ \citenamefont
  {Feldmeier}(2016)}]{Neff:2016ajx}%
  \BibitemOpen
  \bibfield  {author} {\bibinfo {author} {\bibfnamefont {T.}~\bibnamefont
  {Neff}}\ and\ \bibinfo {author} {\bibfnamefont {H.}~\bibnamefont
  {Feldmeier}},\ }\href@noop {} {\  (\bibinfo {year} {2016})},\ \Eprint
  {http://arxiv.org/abs/1610.04066} {arXiv:1610.04066 [nucl-th]} \BibitemShut
  {NoStop}%
\end{thebibliography}%

\end{document}